\documentclass[longauth]{aa}
\usepackage{graphicx}
\usepackage{natbib}

\usepackage{txfonts}
\usepackage{color}

\newcommand{\kms}{\mbox{$\mbox{km\,s}^{-1}$}\,}

\begin{document}

\title{HARPS-N high spectral resolution observations of Cepheids \\ II. The impact of the surface-brightness color relation on the Baade-Wesselink projection factor of $\eta$~Aql 
}
\titlerunning{HARPS-N observations of $\eta$~Aql}
\authorrunning{Nardetto et al. }

\author{N.~Nardetto \inst{1} 
\and W.~Gieren\inst{2,3}  
\and J.~Storm \inst{4}  
\and V.~Hocd\'e \inst{5} 
\and G.~Pietrzy\'nski \inst{5}  
\and P.~Kervella \inst{6}  
\and A.~M\'erand \inst{7} 
\and A.~Gallenne\inst{2,8}  
\and D.~Graczyk \inst{5} 
\and B.~Pilecki \inst{5} 
\and E.~Poretti \inst{9}
\and M.~Rainer \inst{9} 
\and B. Zgirski \inst{5}
\and P. Wielg\'orski \inst{5}  
\and G. Hajdu \inst{5}  
\and M. G\'orski \inst{5}  
\and P. Karczmarek \inst{2}  
\and W. Narloch \inst{2}  
\and M. Taormina \inst{5}  
}
\institute{Universit\'e C\^ote d'Azur, OCA, CNRS, Lagrange, France,  Nicolas.Nardetto@oca.eu   
\and Universidad de Concepci\'on, Departamento de Astronom\'ia, Casilla 160-C, Concepci\'on, Chile
\and Millenium Institute of Astrophysics, Santiago, Chile 
\and Leibniz Institute for Astrophysics, An der Sternwarte 16, 14482, Potsdam, Germany 
\and Nicolaus Copernicus Astronomical Center, Polish Academy of Sciences, ul. Bartycka 18, PL-00-716 Warszawa, Poland
\and LESIA (UMR 8109), Observatoire de Paris, PSL, CNRS, UPMC, Univ. Paris-Diderot, 5 place Jules Janssen, 92195 Meudon, France 
\and European Southern Observatory, Alonso de C\'ordova 3107, Casilla 19001, Santiago 19, Chile 
\and Unidad Mixta Internacional Franco-Chilena de Astronom\'ia (CNRS UMI 3386), Departamento de Astronom\'ia, Universidad de Chile, Camino El Observatorio 1515, Las Condes, Santiago, Chile
\and  INAF -- Osservatorio Astronomico di Brera, Via E. Bianchi 46, 23807 Merate (LC), Italy 
}

\date{Received ... ; accepted ...}

\abstract{The Baade-Wesselink (BW) method of distance determination of Cepheids is used to calibrate the distance scale. Various versions of this method are mainly based on interferometry and/or the surface-brightness color relation (SBCR).}
{We quantify the impact of the SBCR, its slope, and its zeropoint on the projection factor.  This quantity is used to convert the pulsation velocity into the radial velocity in the BW method. We also study the impact of extinction and of  a potential circumstellar environment on the projection factor.} 
{ We analyzed HARPS-N spectra of $\eta$~Aql to derive its radial velocity curve using different methods. We then applied the inverse BW method using various SBCRs in the literature in order to derive the BW projection factor. } 
{We find that the choice of the SBCR is critical: a scatter of about 8\% is found in the projection factor for different SBCRs in the literature.  The uncertainty on the coefficients of the SBCR affects the statistical precision of the projection factor only little (1-2\%). Confirming previous studies, we find that the method with which the radial velocity curve is derived is also critical, with a potential difference on the projection factor of 9\%. 
An increase of 0.1 in $ E(B-V)$ translates into a decrease in the projection factor of 3\%. A 0.1 magnitude effect of a circumstellar envelope (CSE) in the visible domain is rather small on the projection factor, about 1.5\%. However, we find that a $0.1$ mag infrared excess in the $K$ band due to a CSE can increase the projection factor by about 6\%.} 
{The impact of the surface-brightness color relation on the BW projection factor is found to be critical. Efforts should be devoted in the future to improve the SBCR of Cepheids empirically, but also theoretically, taking their CSE into account as well.}
\maketitle

\section{Introduction}\label{s_Introduction}


The surface brightness color relation (SBCR) is a fundamental tool in modern astronomy. It is indeed used to easily and directly determine the angular diameter of any star from its photometric measurements, usually in two bands, in the visible and the near-infrared. The SBCRs are used, for instance, to derive the distances of eclipsing binaries in the Large Magellanic Cloud \citep{gp13, gallenne18, gp19} and in the Small Magellanic Cloud \citep{graczyk20} with exquisite precision. The reverse is also possible: if the parallax of a star (e.g., from the Gaia mission) is precise enough, the SBCR can be calibrated from eclipsing binaries in the Milky Way \citep{graczyk21}. The SBCRs are usually calibrated from interferometric observations, however \citep{dibenedetto05, kervella04, salsi20, salsi21}. Applying a homogeneous approach on a large number of interferometric data, these recent studies have shown that the SBCRs depend not only on the temperature of stars, but also on their luminosity class. This result was also confirmed from atmospheric models \citep{salsi22}. The SBCRs are also of high interest for the study of transiting exoplanet host stars. The radius of the planet can be directly derived from the radius of the star through the transit \citep{ligi16}. As an example, the SBCRs are currently implemented in the pipeline of the PLAnetary Transits and Oscillation of stars (PLATO) space mission, which provides an independent estimate of the stellar radius \citep{gent22}. The Stellar Parameters and Images with a Cophased Array (SPICA) interferometer installed at the Center for High Angular Resolution Astronomy (CHARA) will play an important role in this respect in the near future \citep{mourard18, pannetier20, mourard22}.

Historically, the SBCRs are also used in the Baade-Wesselink (BW) method of Cepheid distance determination. Since their period-luminosity (PL) relation was established \citep{leavitt1912}, Cepheid variable stars have been used to calibrate the distance scale \citep{hertzsprung13} and then the Hubble constant \citep{riess11, freedman12, riess16}. The BW method was first described by \cite{lindermann18} and was later extended by \cite{baade26} and \cite{wesselink46}. For almost a century, the BW method was used to derive the distance of Cepheids.  The concept is simple: Distances are computed using measurements of the angular diameter over the whole pulsation period along with the stellar radius variations deduced from the integration of the pulsation velocity $V_{\rm puls}$. The latter is linked to the observed radial velocity (RV)
by the projection factor $p=V_{\rm puls}/RV$  \citep{hindsley86, nardetto04, nardetto09}. The three basic versions of  the BW method correspond to different ways of determining the angular diameter curve: the photometric  version based on the SBCRs  \citep{fouque97, fouque07, storm11a, storm11b}, the 
interferometric version \citep{lane00, kervella04a, merand05}, and the more recent version that combines several photometric bands, velocimetry. and interferometry \citep[SPIPS; ][]{merand15}. Recently, a study has shown that the projection factors of Cepheids are highly dispersed (even for Cepheids with the same period), which limits the precision of the BW method to 5-10\% \citep{trahin21}. 

For the long-period Cepheid $\ell$~Car, the angular diameter curves derived from the infrared surface brightness relation and infrared interferometry, respectively, are consistent \citep{kervella04d}, while this is not the case for the short-period Cepheid $\delta$~Cep \citep{ngeow12}. In the SPIPS approach, \citet{merand15} have resolved this discrepancy by adding an ad hoc{\it } infrared excess, which can be justified by the presence of a circumstellar environment (CSE) around Cepheids. The CSE of Cepheids was discovered by interferometry \citep{kervella06a, merand06, merand07, gallenne13b, hocde21}. Interestingly, a resolved structure around $\delta$ Cep was discovered in the visible spectral range using interferometry \citep{nardetto16a}. Recently, \citet{hocde20a} showed that the infrared excess of the Cepheids can be explained by a shell of ionized gas, whereas the favored hypothesis until now was dust. At the same time, \citet{hocde20b} also showed that the ionization of the gas could come from shocks propagating in the chromosphere of the Cepheids. 
It appears from these studies that the CSE, like the projection factor, might bias the BW distance.  

We secured spectroscopic observations with the High Accuracy Radial velocity Planet Searcher for the Northern hemisphere (HARPS-N) of the short period Cepheids $\delta$ Cep and $\eta$~Aql. \citet{nardetto17} analyzed the projection factor of $\delta$ Cep. In this study, we focus on the SBCR issue. The overall goal is to clarify the impact of the SBCR on the BW projection factor. Indeed, many SBCRs are described in the literature, but their coefficients differ significantly \citep{salsi20}. We start in Sect.~\ref{s_HARPS-N} with the presentation of the HARPS-N data of $\eta$~Aql, from which we derive a radial velocity curve as well as the atmospheric velocity gradient. Then, we collect photometric data (Sect.~\ref{s_photo}) and combine them with the SBCRs available in the literature (Sect.~\ref{s_SBCR}) in order to apply the inverse BW method and derive the projection factors (Sect. ~\ref{s_BW}). We then further explore the impact of various parameters on the projection factor: zeropoint and slope of the SBCR, $ E(B-V),$  and a possible CSE (Sect.~\ref{s_impact}). We conclude in Sect.~\ref{s_conclusion}. This paper is part of the international Araucaria~Project, whose purpose is to provide an improved local calibration of the extragalactic distance scale out to distances of a few megaparsecs~\citep{gieren05_messenger}.

\section{HARPS-N spectroscopic observations}\label{s_HARPS-N}

HARPS-N is a high-precision radial-velocity spectrograph installed at the Italian Telescopio Nazionale Galileo (TNG), which is a 3.58-meter telescope located at the Roque de los Muchachos Observatory on the island of La Palma, Canary Islands, Spain \citep{co12}.  HARPS-N is the northern hemisphere counterpart of the similar HARPS instrument installed at the ESO 3.6 m telescope at La Silla Observatory in Chile. The instrument covers the wavelength range from 3800 to 6900~$\,$Angstrom with a resolving power of $R \simeq 115000$. A total of 98 spectra were secured between 27 March and 8 September 2015 in the framework of the OPTICON proposal 2015B/015. In order to calculate the pulsation phase of each spectrum, we used the recent ephemeris $P=7.1765470$~d, $T_0=2411998.2930$~d, and  $\frac{dP}{dt}=0.00000295060$ days/year from \citet{Cso22}.
The data are spread over 13 of the 23 pulsation cycles  that elapsed between the first and last epoch.
The final products of the HARPS-N data reduction software (DRS) installed at TNG (online mode)
are background-subtracted, cosmic-corrected, flat-fielded, and wavelength-calibrated spectra
(with and without merging of the spectral orders).

\begin{figure}[htbp]
\begin{center}
\resizebox{0.8\hsize}{!}{\includegraphics[clip=true]{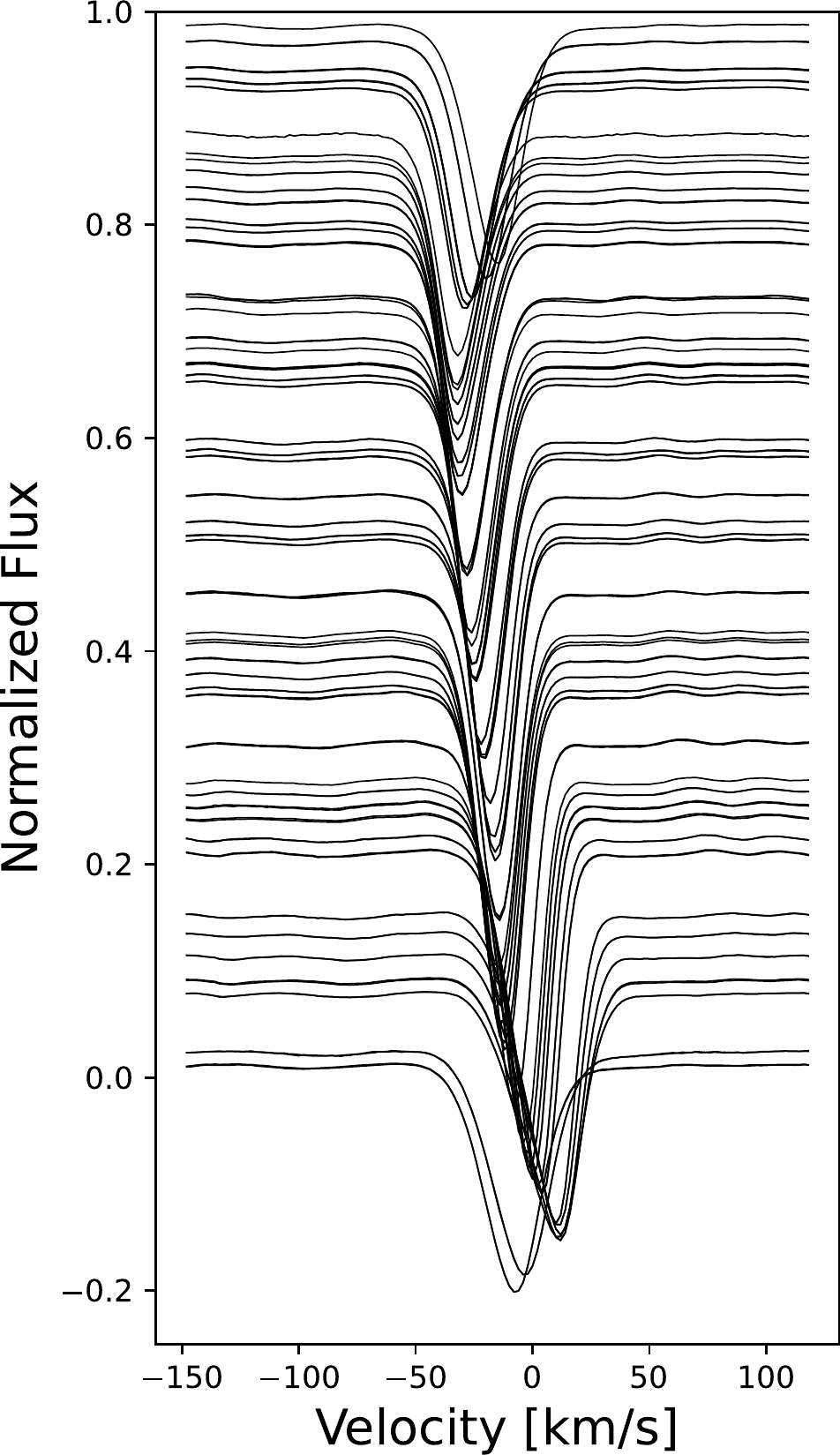}}
\end{center}
\caption{Cross-correlated spectral line profile (using a template for an F6 star) as a function of the pulsation phase (indicated on the left of the diagram).}
\label{fig_prof}
\end{figure}

The DRS computes the cross-correlation function (noted 'cc' in the following) using a mask including thousands of lines covering the whole HARPS-N spectral ranges. The observer can  select the mask among the masks that are available online, and the G2V mask was the closest to the $\eta$~Aql spectral type (F6I). As a further step, we recomputed the cross-correlation function by using the HARPS-N DRS in the offline mode on the Yabi platform, considering a custom mask for a F6I star. Yabi \citep{hunter12} is a Python web application installed at IA2\footnote{ \url{https://www.ia2.inaf.it}} in Trieste that allows authorized users to run the HARPS-N DRS pipeline on their own proprietary data with custom input parameters.
  Fig.~\ref{fig_prof} shows that the mean profiles reflect the large-amplitude radial pulsation. Then, the DRS computes the stellar radial velocity by fitting a Gaussian to the cross-correlation functions ($RV_\mathrm{cc-g}$).  However, \citet{nardetto06a} have shown that the centroid velocity, that is, the first moment (or centroid) of the spectral line profile, is estimated as 

\begin{equation} \label{Eq_CDG}
 RV_{\mathrm c} =  \frac{c}{\lambda_\mathrm{0}}\frac{\int_{\rm line} (\lambda - \lambda_\mathrm{0}) S(\lambda) d\lambda}{\int_{\rm line} S(\lambda) d\lambda}
,\end{equation}
where $S(\lambda)$ is the observed line profile, $c$ is the velocity of light, and $\lambda_\mathrm{0}$ is the rest wavelength of the spectral line. This definition, at least on single lines, is most frequently adapted in the context of the BW method because this velocity is independent of stellar rotation or spectral line width variation. By applying the first moment to the cross-correlation function, we can derive the $RV_\mathrm{cc-c}$ velocity. 



\subsection{Cross-correlated radial velocity curves}

\begin{figure}[htbp]\begin{center}
\resizebox{0.83\hsize}{!}{\includegraphics[clip=true]{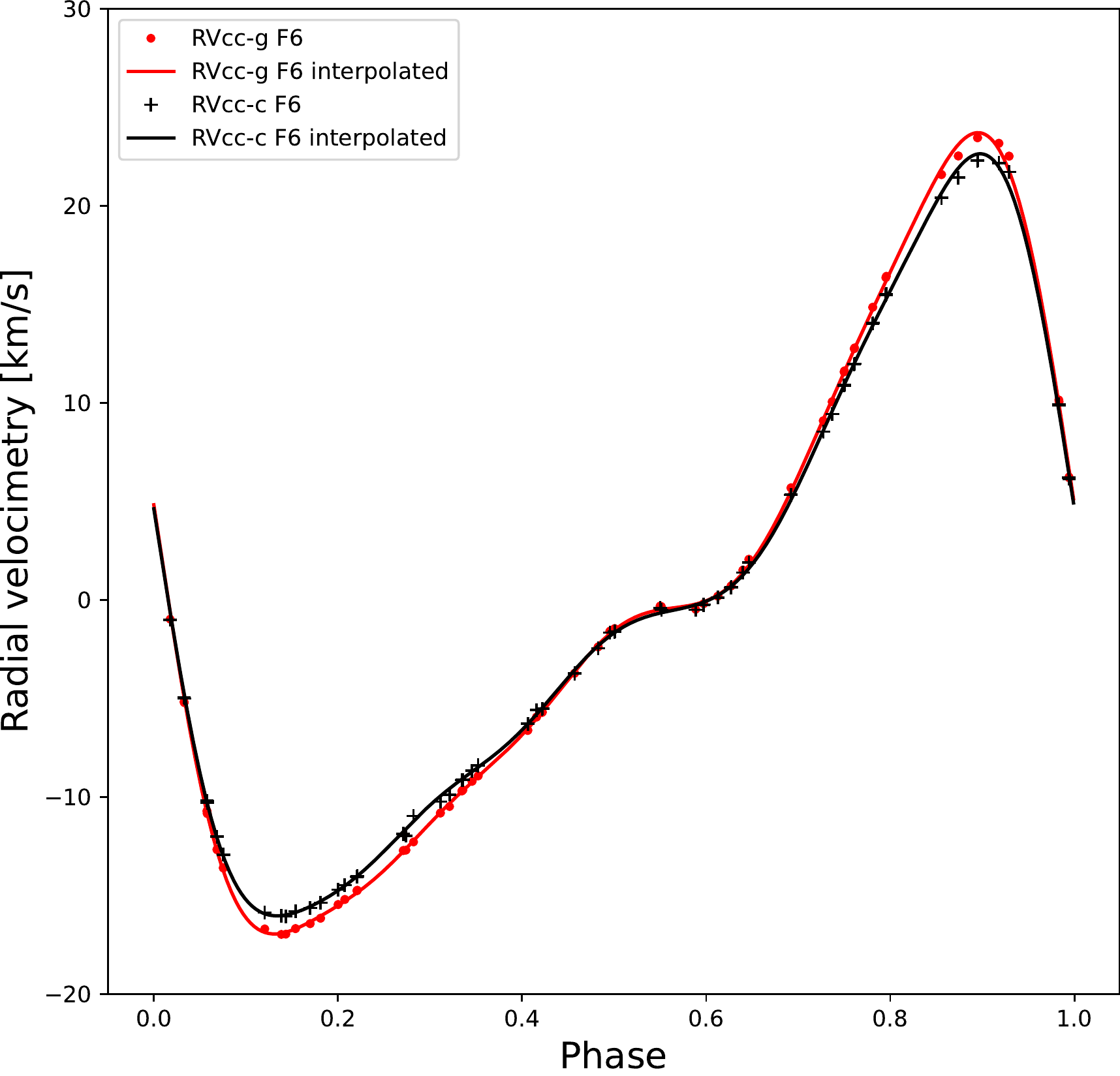}}
\resizebox{0.83\hsize}{!}{\includegraphics[clip=true]{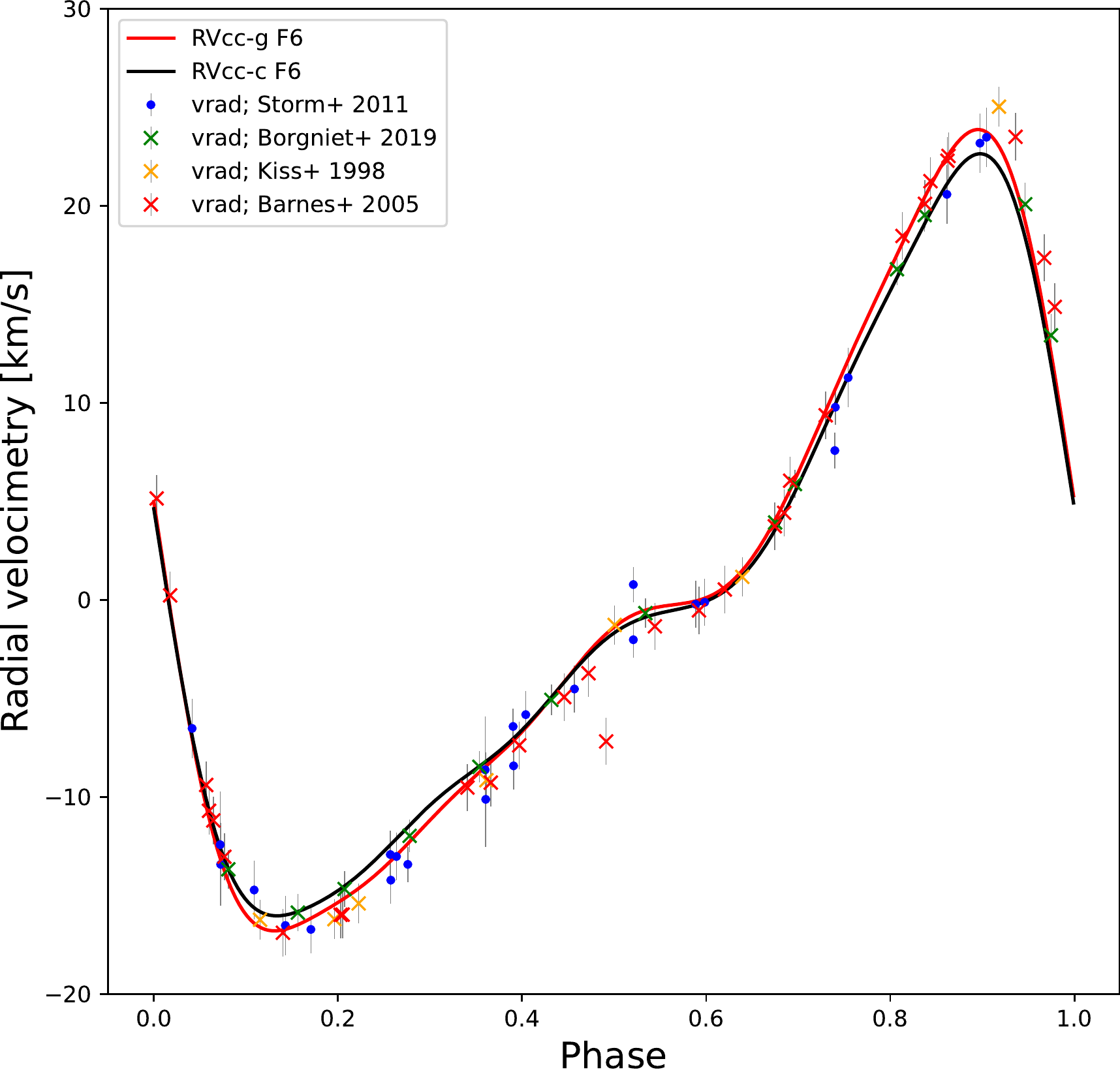}}
\end{center}
\vspace{-0.3cm}
\caption{Radial velocity curves of $\eta$~Aql. Top: HARPS-N cross-correlated measurements (F6 template) using the Gaussian fit (RV$_\mathrm{cc-g}$, red points) and the centroid (RV$_\mathrm{cc-c}$, black points) methods to measure the velocity together with the corresponding Fourier-interpolated curves. Bottom: HARPS-N interpolated curves are shown together with the RV$_\mathrm{cc-g}$ measurements in the literature.}
\label{fig_RVcurves1}
\end{figure}

We obtained four RV curves: $\mathrm{RV_\mathrm{cc-g}}$  and $\mathrm{RV_\mathrm{cc-c}}$ , combined with masks for spectral types G2V and F6I. We find no difference between G2 and F6I templates. We decided to keep the F6 template in the following. Table \ref{Tab_log_ccgF6} and Table \ref{Tab_log_cccF6} list the $\mathrm{RV_\mathrm{cc-g}}$ and $\mathrm{RV_\mathrm{cc-c}}$ values obtained with the custom F6I mask, respectively, while the data and the Fourier-interpolated radial velocity curves are shown in the top panel of Fig.~\ref{fig_RVcurves1}.  Measurements and interpolated curves in the figure were corrected from a $\gamma$-velocity of $14.87$~\kms corresponding to the average of the interpolated radial velocity curve $\mathrm{RV_\mathrm{cc-c}}$ (F6 mask). In the bottom panel of Fig.~\ref{fig_RVcurves1}, we overplot the $\mathrm{RV_\mathrm{cc-g}}$ and $\mathrm{RV_\mathrm{cc-c}}$ interpolated curves of HARPS-N with  previous measurements found in the literature \citep{storm11a, borgniet19, kiss98, barnes05}. The agreement between $\mathrm{RV_\mathrm{cc-g}}$ and previous measurements (obtained with the same method) is good.

We find no evidence for cycle-to-cycle differences in the RV amplitude as exhibited by long-period Cepheids \citep{anderson14,anderson16}. We find a significant difference between the $\mathrm{RV_\mathrm{cc-g}}$ and $\mathrm{RV_\mathrm{cc-c}}$ curves in terms of amplitude. The impact of this difference on the projection factor is discussed in Sect. \ref{s_impact}.





\begin{figure}[htbp]\begin{center}
\resizebox{0.83\hsize}{!}{\includegraphics[clip=true]{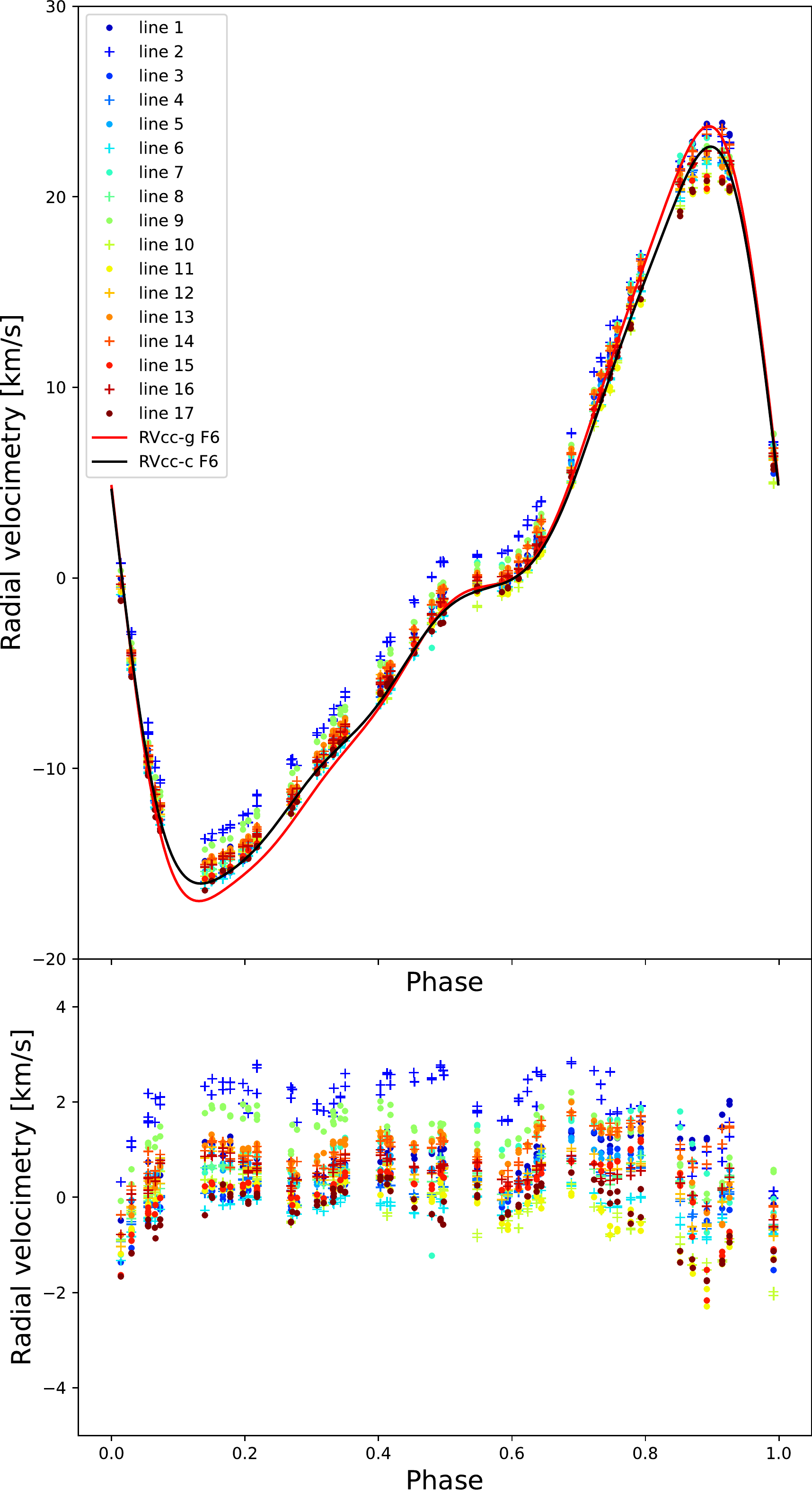}}
\end{center}
\vspace{-0.3cm}
\caption{Radial velocity curves of $\eta$~Aql for different spectral lines.Top: HARPS-N interpolated curves are shown together with the radial velocity derived using the centroid method RV$_\mathrm{cc-c}$ for the 17 unblended spectral lines listed in Table~\ref{tab_lines}. Bottom: Difference compared to the RV$_\mathrm{cc-c}$ curve. }
\label{fig_RVcurves2}
\end{figure}


\begin{table}
\begin{center}
\caption{Spectral lines with their wavelength of reference $\lambda_\mathrm{0}$, excitation potential (Ep), and oscillator strengths  $\log(gf)$. }\label{tab_lines}
\begin{tabular}{llcccc}
\hline \hline \noalign{\smallskip}
Number & line   &   $\lambda_\mathrm{0}$    & Ep & $\log(gf)$ \\ 
\hline
1 & \ion{Fe}{I} &       4683.560        &       2.831   &       $-2.319$  \\ 
2 & \ion{Fe}{I} &       4896.439        &       3.883   &       $-2.050$  \\ 
3 & \ion{Ni}{I} &       5082.339        &       3.658   &       $-0.540$  \\ 
4 & \ion{Fe}{I} &       5367.467        &       4.415   &       $0.443$   \\ 
5 & \ion{Fe}{I} &       5373.709        &       4.473   &       $-0.860$  \\ 
6 & \ion{Fe}{I} &       5383.369        &       4.312   &       $0.645$   \\ 
7 & \ion{Ti}{II}        &       5418.751        &       1.582   &       $-2.110$  \\ 
8 & \ion{Fe}{I} &       5576.089        &       3.430   &       $-1.000$  \\ 
9 & \ion{Fe}{I} &       5862.353        &       4.549   &       $-0.058$  \\ 
10 & \ion{Fe}{I}        &       6003.012        &       3.881   &       $-1.120$  \\ 
11 & \ion{Fe}{I}        &       6024.058        &       4.548   &       $-0.120$  \\ 
12 & \ion{Fe}{I}        &       6027.051        &       4.076   &       $-1.089$  \\ 
13 & \ion{Fe}{I}        &       6056.005        &       4.733   &       $-0.460$  \\ 
14 & \ion{Si}{I}        &       6155.134        &       5.619   &       $-0.400$  \\ 
15 & \ion{Fe}{I}        &       6252.555        &       2.404   &       $-1.687$  \\ 
16 & \ion{Fe}{I}        &       6265.134        &       2.176   &       $-2.550$  \\ 
17 & \ion{Fe}{I}        &       6336.824        &       3.686   &       $-0.856$  \\  
\hline
     &    & \AA &  eV & \\

\hline \noalign{\smallskip}
\end{tabular}
\end{center}
\end{table}


\begin{figure*}[htbp]
\begin{center}
\centering
        \includegraphics[scale=0.5]{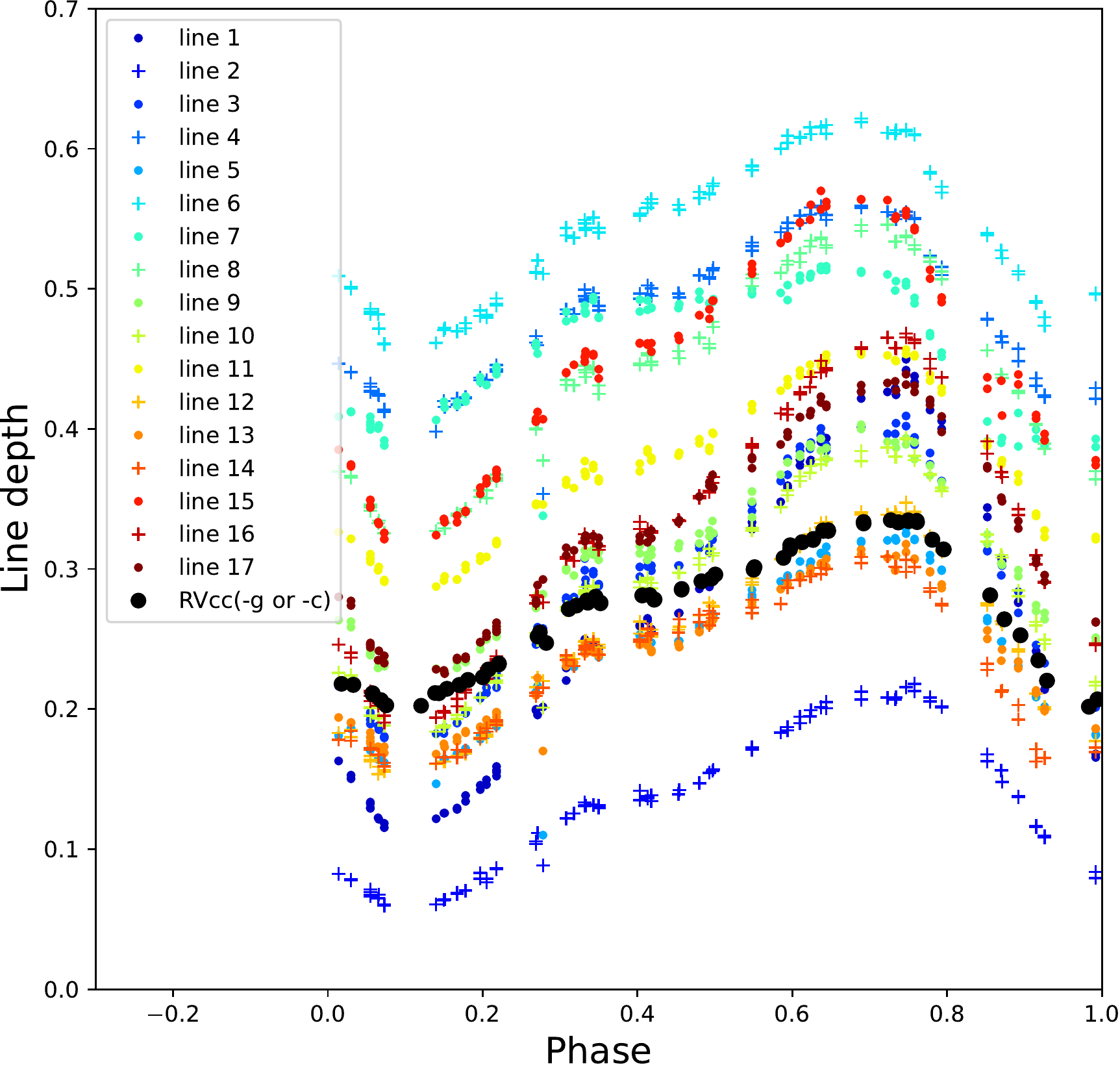}
        \includegraphics[scale=0.5]{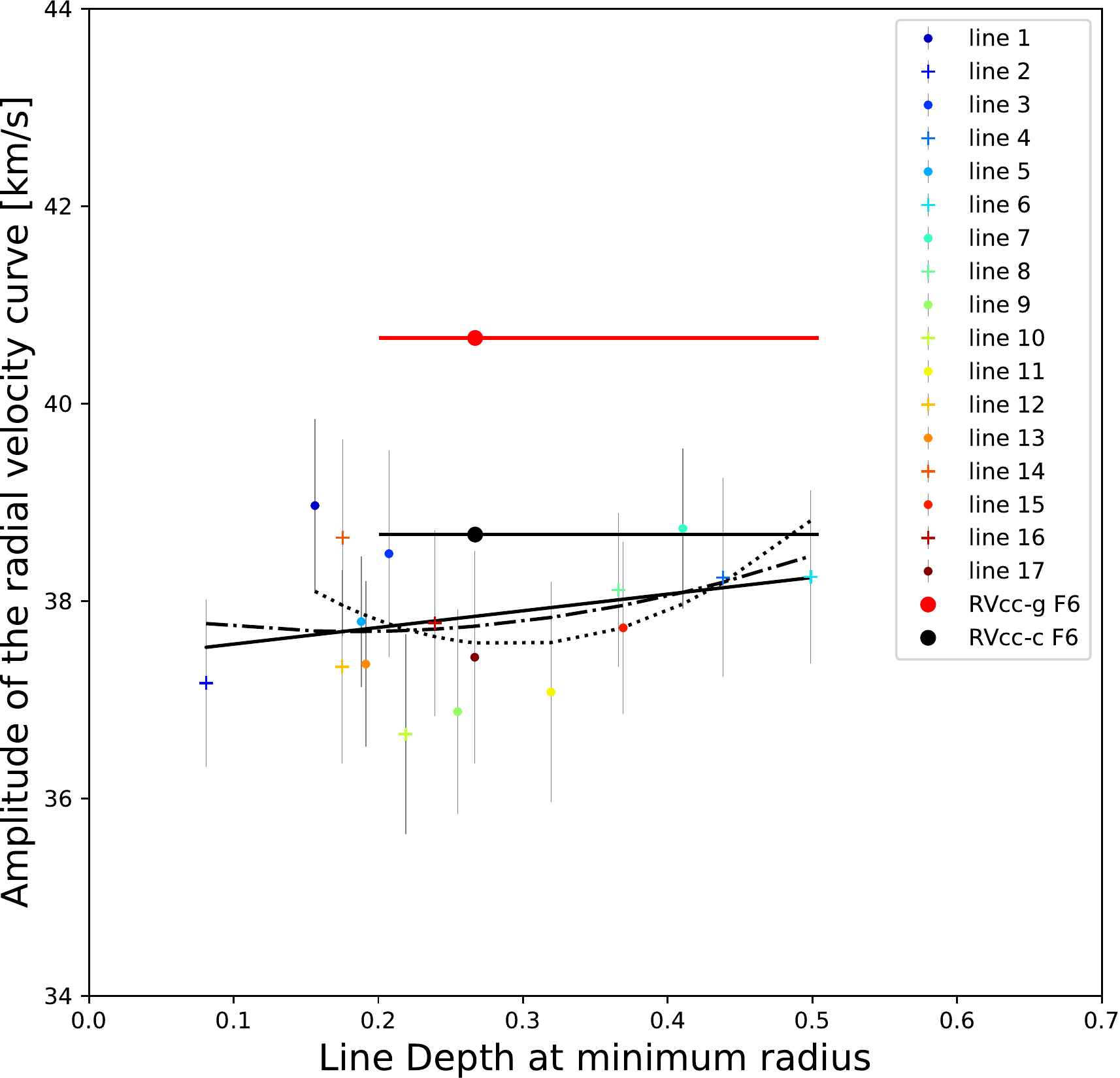}
        \hfill

\end{center}
\caption{Measure of the atmospheric velocity gradient of $\eta$~Aql. Left: Depth for the 17 spectral lines as a function of the pulsation phase, as well as the depth of the cross-correlated profile (black dots). The quantity corresponding to the line depth is slightly different for the Gaussian fit (RV$_\mathrm{cc-g}$) method or the centroid (RV$_\mathrm{cc-c}$), but it provides the same values. Right:  Amplitude of the radial velocity curves of the 17 spectral lines listed in Table~\ref{tab_lines},  plotted vs. the line depth, at minimum radius. The amplitudes for the RV$_\mathrm{cc-g}$ and RV$_\mathrm{cc-c}$ curves (template F6) are shown for comparison (red and black dots, respectively) together with the minimum and maximum depth of the cross-correlated profile (horizontal red and black lines). Different fits are processed (see the text for the explanation).} \label{Fig_grad}
\end{figure*}

\subsection{Atmospheric velocity gradient}\label{s_p}

In this section, we apply the same approach as was presented in \citet{nardetto17}. In brief, we considered the 17 unblended spectral lines presented in Table~\ref{tab_lines}. For each of these lines, we derived the centroid velocity ($RV_{\mathrm c}$) as defined previously (Eq. \ref{Eq_CDG}). The radial velocity measurements associated with the spectral lines are presented in Fig.~\ref{fig_RVcurves2}. The plotted RV$_\mathrm{c}$ curves were corrected for the $\gamma$-velocity value corresponding to the RV$_\mathrm{cc-c}$ curve. The residuals, that is, the $\gamma$-velocity offsets, between the curves are related to the line asymmetry and the k-term value (see \citealp{nardetto08a} for Cepheids and \citealp{nardetto13, nardetto14} for other types of pulsating stars). \citet{nardetto07} split the projection factor into three quantities: $p= p_{\mathrm{o}}\,f_{\mathrm{grad}}\,f_{\mathrm{o-g}}$, where $p_\mathrm{0}$ is the geometrical projection factor (linked to the limb darkening of the star); $f_\mathrm{grad}$, which  is a cycle-integrated quantity linked to the velocity gradient in the atmosphere of the star (i.e., between the considered line-forming region and the photosphere); and $f_\mathrm{o-g}$, which is the relative motion of the optical pulsating photosphere with respect to the corresponding mass elements.

We derived $f_{\mathrm{grad}}$ associated with line $i$ directly from HARPS-N observations using
\begin{equation} \label{Eq_grad2}
 f_{\mathrm{grad}}(i)= \frac{b_0}{\Delta RV_{\mathrm{c}} (i)}, 
\end{equation}
where
\begin{equation} \label{Eq_grad}
\Delta RV_{\mathrm{c}} (i) = a_0 D (i) + b_0, 
\end{equation}
with $ \Delta RV_{\mathrm{c}}(i)$ the amplitude of the radial velocity curves associated with line $i$ and $D(i)$, the line depth at minimum radius. In this definition, $b_0$ is the amplitude of the radial velocity associated with the photosphere of the star. By integrating RV$_\mathrm{cc-c}$ , we find that the phase corresponding to the minimum radius is $\phi=0.015$.  In the left panel of Fig.~\ref{Fig_grad}, we plot the line depth as a function of the pulsation phase associated with each of the 17 lines, as well as the depth of the cross-correlated profile. The RV$_\mathrm{c}$  and the line depth curves associated with each line were then interpolated using a Fourier analysis. In the right panel of Fig.~\ref{Fig_grad}, $\Delta RV_{\mathrm{c}}$ is plotted as a function of $D$ for each individual line. The values obtained for the RV$_\mathrm{cc-g}$ and RV$_\mathrm{cc-c}$ curves are also shown for comparison, as well as the whole range of the associated line depth over the cycle. We fit a linear trend to the data. We find 

\begin{equation} \label{Eq_grad_O}
\Delta RV_{\mathrm{c}}= [1.7 \pm 1.9] D +  [37.4 \pm 0.5] 
.\end{equation}
The reduced $\chi^2$ is 0.48. When a quadratic curve was fit that include line 2 (dash-dotted line) or excluded line 2 (dotted line), we obtained a reduced $\chi^2$ of 0.50 and 0.43, respectively. In any case, the results are consistent with a null atmospheric velocity gradient for $\eta$ Aql ($\Delta RV_{\mathrm{c}}=37.89 \pm 0.23,$ with a reduced $\chi^2$ of 0.49). 
 However,  the differences in the velocity amplitudes found for the 17 lines are large enough to affect the projection significantly, as discussed in Sect. \ref{s_impact}. 

\section{Photometric data}\label{s_photo}

\begin{figure}[htbp]
\begin{center}
\resizebox{0.9\hsize}{!}{\includegraphics[clip=true]{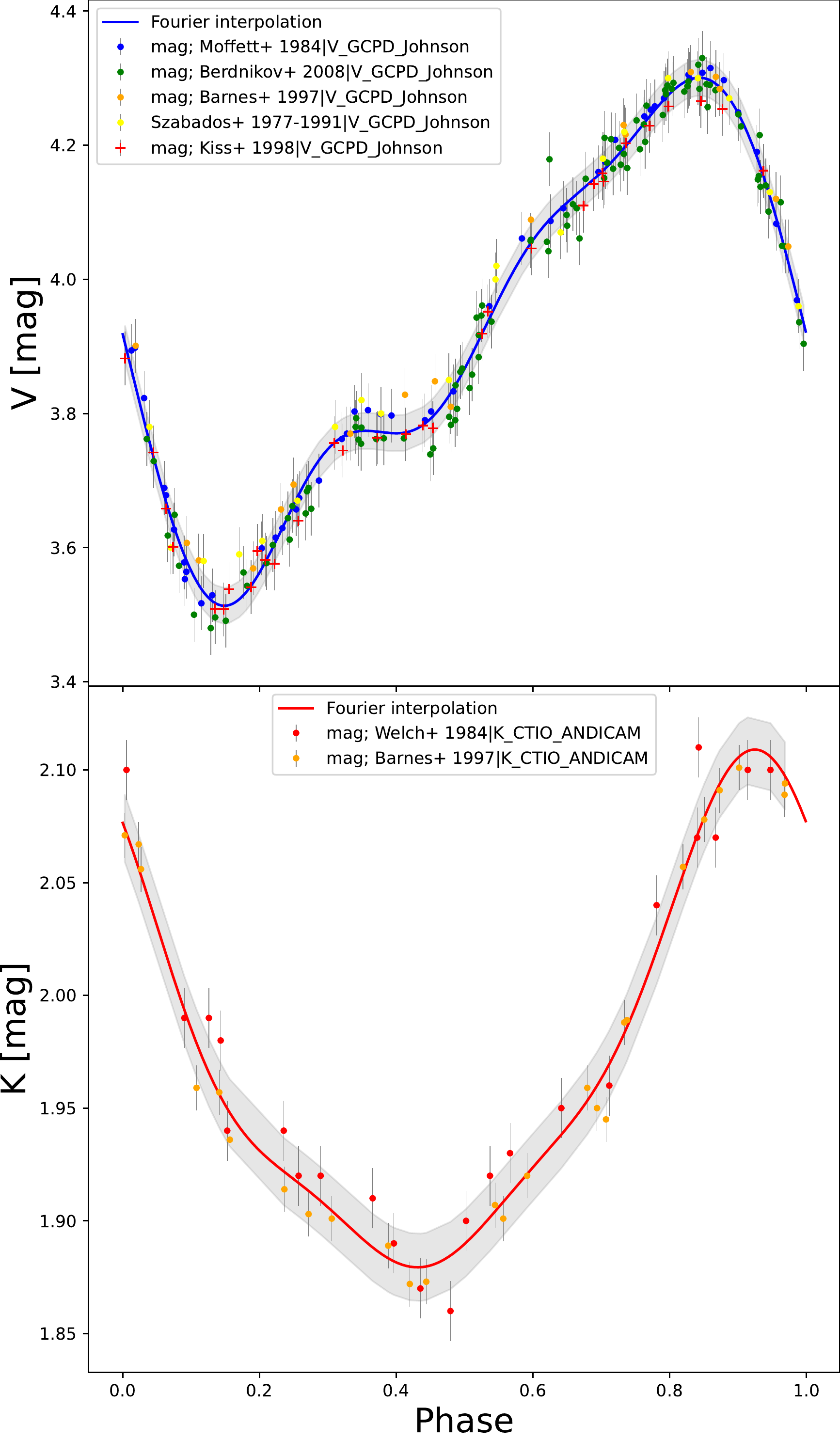}}
\end{center}
\caption{Photometric curves of $\eta$~Aql. Top: V-band photometric measurements from the literature plotted against the pulsation phase. The solid blue line is an interpolation using a Fourier series. Bottom: Same as in the top panel, but in the K-band.  The gray zones correspond to the uncertainty in the fit. }\label{fig_photo}
\end{figure}

In order to apply the SBCR, we need photometry in the visual and in the infrared domains. For the visual photometry, we used data from
\citet{moffett84}, \citet{berdnikov08}, \citet{barnes97}, \citet{szabados77} and \citet{kiss98}, while for infrared photometry, we considered
\citet{welch84} and \citet{barnes97}. The curves were interpolated using a Fourier analysis, and they are plotted in Fig.\ref{fig_photo}. The visual photometric data are in the Johnson system, while the infrared data are in the Cerro Tololo Inter-American Observatory (CTIO) system \citep{elias82}.

\section{Comparison of the surface brightness color relations}\label{s_SBCR}

As already shown previously \citep{dibenedetto93}, but using a method based on various selection criteria, \citet{salsi20, salsi21} have again confirmed the result that the SBCR depends on the stellar class, that is, on the stellar surface gravity. The difference is significant between dwarfs and giants, but it remains unclear between luminosity classes III, II, and~I. Thus, when the SBCR version of the BW method is to be applied to Cepheids, it is important to consider an SBCR that is suitable for Cepheids. In Table~\ref{tab_lin_sbcr} and Table~\ref{tab_nlin_sbcr}  we list the linear and nonlinear SBCRs from the literature that might be applicable to Cepheids, respectively. Some of the SBCRs are specific to Cepheids, that is, they are based on observations of Cepheids (indicated by "Cep" in Tables~\ref{tab_lin_sbcr} and~\ref{tab_nlin_sbcr}), while others are typically for stellar classes III, II, or I. Some of the SBCRs were considered as valid for all classes by their authors. They are indicated by "all" in the tables. The photometric systems used to calibrate the linear and nonlinear SBCRs are provided in Tab.~\ref{tab_systems}. The systems used are the Johnson system \citep{johnson66}, the CTIO system \citep{elias82}, the South African Astronomical Observatory (SAAO) system \citep{carter90}, and the Two-Micron Sky Survey (TMSS) system, also called InfraRed Caltech catalog (IRC) \citep{neugebauer69}. In the visual domain, the generally used photometric system is the Johnson system, except for K04, who used the Cousins system. The vB99 SBCR used the Catalog of Infrared Observations (CIO) from \citet{gezari93} for the infrared data and the General Catalog of Photometric Data (GCPD) for the visual magnitudes \citep{mermilliod97}. These two catalogs provide a compilation of infrared and visual data in various photometric systems, respectively. The impact of these systems on the projection factor is discussed in Sect.~\ref{s_impact}.

\begin{figure}[htbp]
\begin{center}
\resizebox{0.9\hsize}{!}{\includegraphics[clip=true]{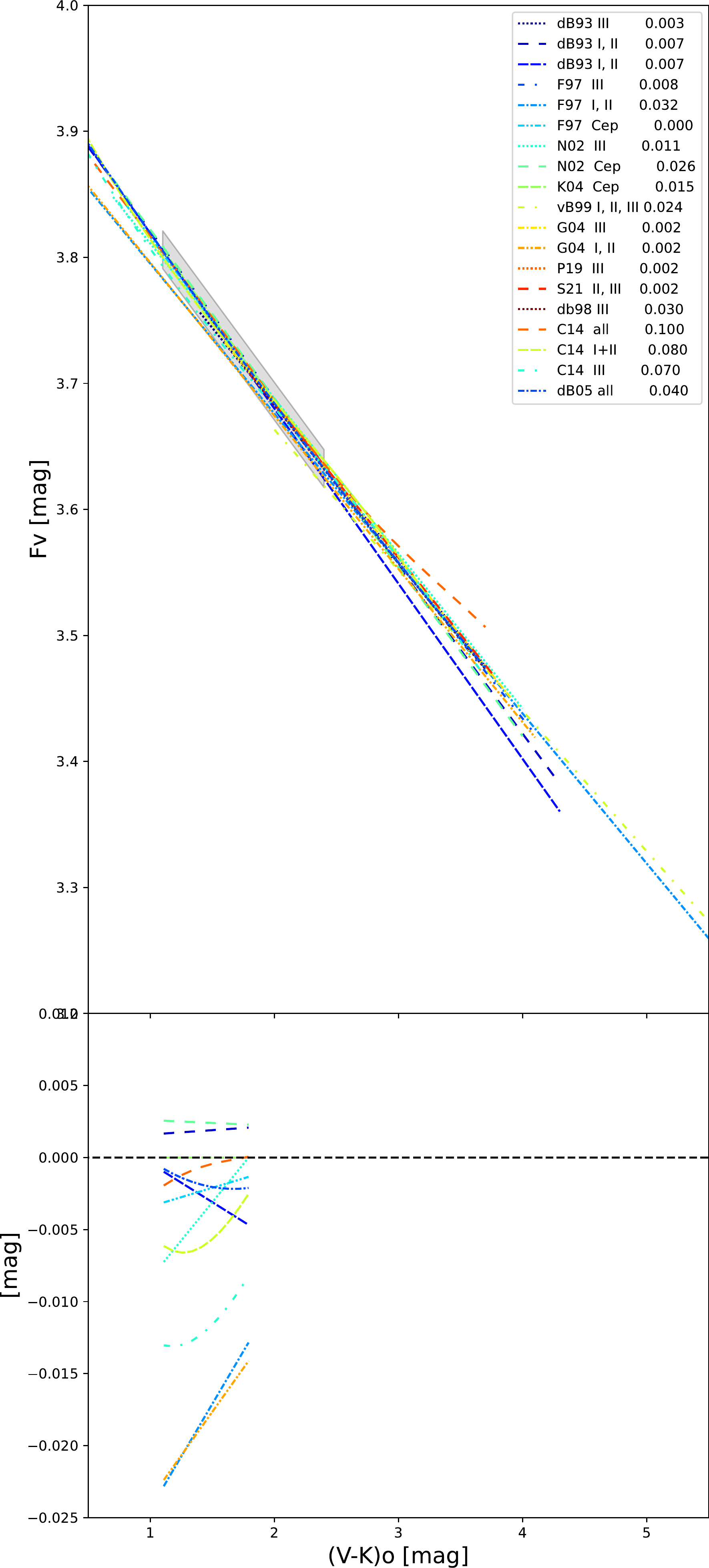}}
\end{center}
\caption{The SBCRs in the litterature that are consistent with an application to Cepheids. Top: SBCR in the literature plotted over the $(V-K)_0$ validity domain. The gray zone correspond to the rms of K04 SBCR, i.e., 0.015 mag. The rms of other SBCRs (in mag) is indicated in the legend. Bottom: SBCRs compared to K04 in the $(V-K)_0$ range corresponding to $\eta$~Aql pulsation. SBCRs that are partly or not at all consistent with the $(V-K)_0$ range of $\eta$ Aql (see Table~\ref{tab_res}) are not indicated in this second panel.}\label{fig_sbcr}
\end{figure}

\begin{figure}[htbp]
\begin{center}
\resizebox{1.0\hsize}{!}{\includegraphics[clip=true]{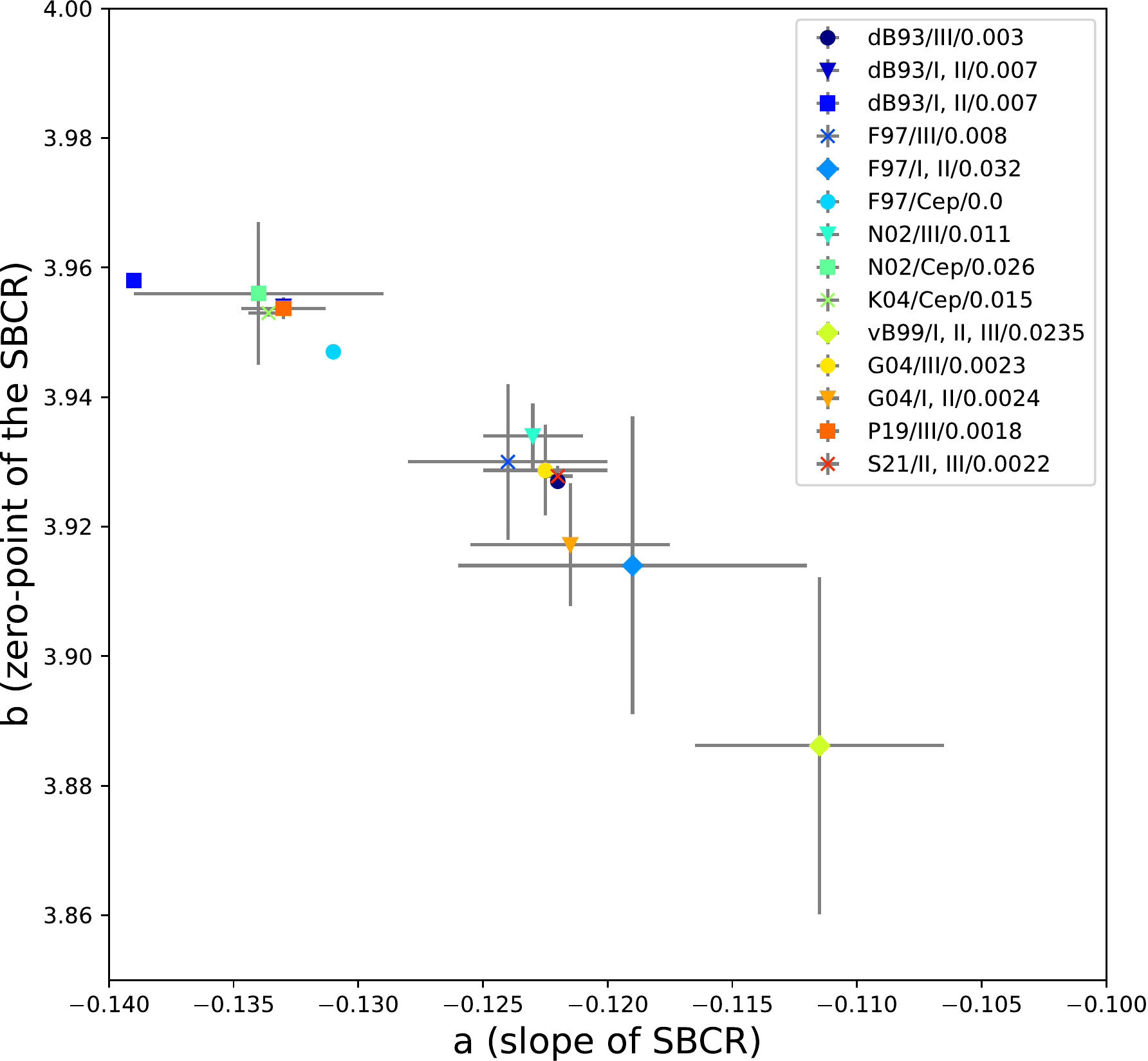}}
\end{center}
\caption{Comparison of the slope and zeropoint of the linear SBCR in the literature.}\label{fig_absbcr}
\end{figure}

In the upper panel of Fig. \ref{fig_sbcr}, we plot the 19 SBCRs of Tables~\ref{tab_lin_sbcr} and~\ref{tab_nlin_sbcr}. In the lower panel, we plot the difference between these SBCRs and the fiducial SBCR of \citet{kervella04c} (K04). In this second panel, we plot only the SBCRs whose validity domain is consistent with the $V-K$ color range of $\eta$~Aql (i.e., from about 1.1 to 1.8 mag). We use the SBCR of K04 as a reference as it is often used in the literature when the BW method is used. As an indication, the rms of the K04 relation is shown in the upper panel of the figure as a gray zone. The rms of the other SBCRs can be found in Tables~\ref{tab_lin_sbcr} and \ref{tab_nlin_sbcr}. The agreement between the 19 SBCRs is better than about 0.0075$\,$mag between 1.5 and 2.5 mag in $V-K$. Outside this range, significant disagreements are found. 
In Fig.~\ref{fig_absbcr} we plot the slope of the linear SBCRs as a function of the zeropoint, with the associated statistical uncertainties when available. The relations disagree clearly, but these disagreements come in a large part from the fact that the slope and zeropoint of the SBCRs are correlated. This is because the barycenter of the measurements is usually not taken as a reference to calculate the zeropoint.

\section{Comparison of the angular diameter curves}\label{s_diameter}

\begin{figure*}[htbp]
\begin{center}
\resizebox{1.0\hsize}{!}{\includegraphics[clip=true]{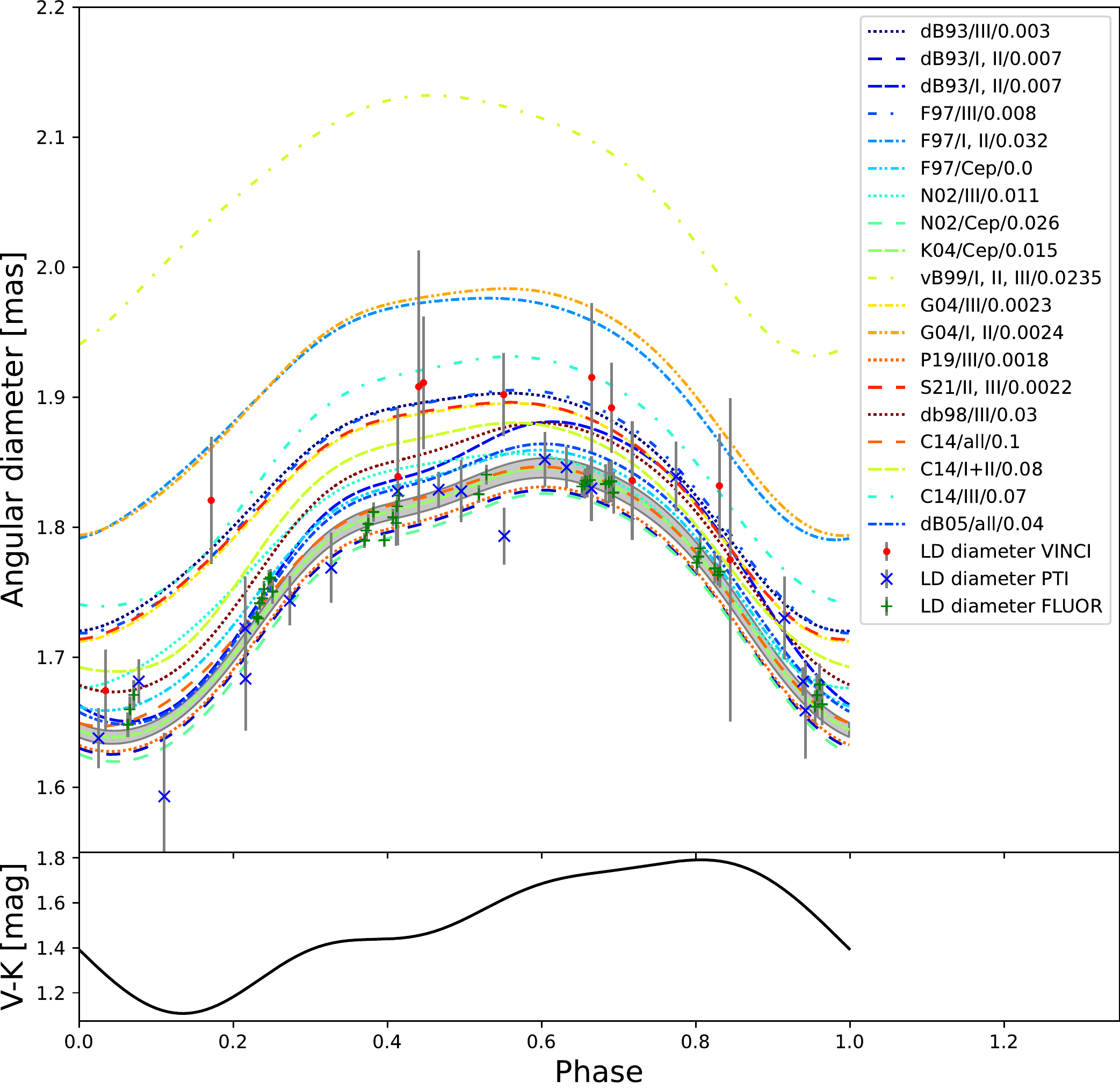}}
\end{center}
\caption{Angular diameter curves as derived from SBCRs. Top: Angular diameter of $\eta$~Aql as a function of the pulsation phase for different SBCRs (from Fig.~\ref{fig_sbcr}) combined with the V and K light curves (Fig.~\ref{fig_photo}).
The gray band corresponds to the uncertainty on the angular diameters caused by the uncertainty of the coefficients of the K04 SBCR. For comparison, we also plot the limb-darkened angular diameter as obtained by VINCI, PTI, and FLUOR measurements. Bottow: $V-K$ color of $\eta$~Aql.}\label{fig_diam}

\end{figure*}

In this section, we apply the SBCRs in order to derive the angular diameter curves. 
To do this, we used the method described in
\citet{fouque97}, \citet{fouque07}, \citet{storm11a} and \citet{storm11b}, which relies on the following relation:
 
\begin{equation}
\label{Eq_Fv}
\log \theta_{\mbox{\scriptsize LD}} (\phi _i) = 8.4414-0.2 V_{0}(\phi_i) - 2 F_{V_{0}}(\phi_i).
\end{equation}

where $\theta_{\mbox{\scriptsize LD}}$, $V,$ and $F_V$ are the limb-darkened angular diameter, the magnitude, and the surface brightness, respectively, in the $V$ band at the corresponding phase of pulsation $\phi_i$. The SBCRs as listed in Table~\ref{tab_lin_sbcr} are defined by
\begin{equation}\label{Eq_lin_sbcr}
F_{V_{0}} = a (V-K)_0 + b
,\end{equation}

where $K$ is the magnitude in the K band. The subscript 0 refers to the magnitudes corrected for interstellar extinction. By definition, the interstellar absorption $A_\mathrm{V}$ in the visible band is given by $A_\mathrm{V} = R_{V} \times E (B-V) $, where $E (B-V)$ is the $B-V$ color excess, and $R_{V}$ is the total-to-selective absorption in the $V$ band.  The nonlinear SBCRs in the literature (Table~\ref{tab_nlin_sbcr}) are most often provided in the form
\begin{equation}
\label{Eq_nlin_sbcr}
S_{V_{0}} = \sum \limits_{k=0}^{k} a_k (V-K)_0^k
.\end{equation}

When $M_\mathrm{bol_\odot}=4.74$ mag and $f_\odot=1361$ W/$m^2$  \citep{prsa16} are used, the conversion into $F_\mathrm{V}$ is by definition

\begin{equation}\label{Eq_conv}
F_\mathrm{V}=4.2196-0.1 S_\mathrm{V}
.\end{equation}

In order to derive the angular diameter curve, we considered the interpolated curves in the V and K bands presented in Sect.~\ref{s_photo}. For the exctinction, we used the value from the \textit{Stilism}\footnote{The online tool is available at \url{http://stilism.obspm.fr}} online tool \citep{Stilism1, Stilism2} to compute the color excess $ E(B-V)$. This tool produces tridimensional maps of the local interstellar matter (ISM). According to this map, the extinction is constant in a certain distance range for $\eta$~Aql. We find $E(B-V) =0.152$ mag in the case of $\eta$~Aql. Using $R_{V}=3.1$ which corresponds to the typical value in the diffuse ISM \citep{cardelli89}, we find  $A_{V}=0.47$ mag and $A_{K} = 0.089 \times A_{V} = 0.04$ mag, according to \citet{reddening}. We arbitrarily set a conservative uncertainty on $A_{V}$ of 0.1 mag. Following \cite{salsi20}, the uncertainty on $A_{K}$ was neglected. The angular diameter values were calculated using Eq.~\ref{Eq_Fv}. Associated with these values, we considered the two following uncertainties \citep{salsi20, salsi20cor}.
The first uncertainty is that on the limb-darkened angular diameter due to the uncertainty on the coefficients of the SBCR, 
\begin{equation}\label{Eq_coe}
\sigma_{\theta_{\mathrm{LD}_\mathrm{coe}}}= 2 \ln(10) \theta_{\mathrm{LD}}  \left\{\left[(V-K)-0.881A_V\right]^2 \sigma_a ^2 + \sigma_b ^2\right\}^{1/2}
.\end{equation}
The second uncertainty is that on the limb-darkened angular diameter that is due to the photometric uncertainties on $V$, $K,$ and $A_V$,
\begin{equation}\label{Eq_pho}
\sigma_{\theta_{\mathrm{LD}_\mathrm{pho}}} = 2 \ln(10)\theta_{\mathrm{LD}} \left\{a^2 \left(\sigma_V^2 + \sigma_K ^2 + 0.014\sigma_{A_{V}}^2\right)\right\}^{1/2}
.\end{equation}

In Fig.~\ref{fig_diam} we plot the angular diameter curves obtained with the 19 SBCRs from Table~\ref{tab_lin_sbcr} and Table~\ref{tab_nlin_sbcr}.
The dark gray zone corresponds to the uncertainty on the coefficients of the SBCR of K04. As a comparison, we overplot the limb-darkened angular diameter derived from interferometry with VINCI/VLTI \citep{kervella04a}, PTI \citep{lane02}, and FLUOR/CHARA \citep{merand15}. The VINCI/VLTI angular diameters are found to be larger (and have larger uncertainties) than the PTI and FLUOR/CHARA measurements, which might be due simultaneously to a lack of data at high spatial frequencies and a bias due to a CSE at low spatial frequencies. In addition, most of the SBCRs fail to simultaneously reproduce the mean and the amplitude of the interferometric (PTI and FLUOR) angular diameter curves.
 We quantify the choice of the SBCR and the interferometric data set on the derived projection factor in Sect. \ref{s_impact}. 





\setlength{\tabcolsep}{0.47cm}
\begin{table*}
\caption{List of all the linear SBCRs found in the literature that are applicable to Cepheids. These SBCRs are provided using the same formalism: $F_{V_{0}} = a  (V-K)_0 + b  $. $l_1=\mathrm{(V-K)}_1$ and $l_2=\mathrm{(V-K)}_2$ are the limit of the validity domain of the SBCRs.}\label{tab_lin_sbcr}
        \begin{center}
                \begin{tabular}{cccccccccc}
                        \hline
                        \hline
ref     &       class   &       $l_1$   &       $l_2$   &       N       &       $b$         &       $\sigma_\mathrm{b}$     &       $a$     &       $\sigma_\mathrm{a}$     &       $\sigma_{rms}$  \\
\hline
dB93    &       III     &       1.4     &       3.7     &       8       &       3.9270  &       0.0000  &       $-0.1220$       &       0.0000  &       0.0030  \\
dB93    &       I, II   &       0.5     &       4.3     &       6       &       3.9540  &       0.0000  &       $-0.1330$       &       0.0000  &       0.0070  \\
dB93    &       I, II   &       0.5     &       4.3     &       6       &       3.9580  &       0.0000  &       $-0.1390$       &       0.0000  &       0.0070  \\
F97     &       III     &       2.22    &       4.11    &       10      &       3.9300  &       0.0120  &       $-0.1240$       &       0.0040  &       0.0080  \\
F97     &       I, II   &       0.52    &       5.53    &       13      &       3.9140  &       0.0230  &       $-0.1190$       &       0.0070  &       0.0320  \\
F97     &       Cep     &       0.8     &       2.4     &       10      &       3.9470  &       0.0000  &       $-0.1310$       &       0.0000  &       0.0000  \\
N02     &       III     &       0.7     &       4       &       57      &       3.9340  &       0.0050  &       $-0.1230$       &       0.0020  &       0.0110  \\
N02     &       Cep     &       0.7     &       4       &       59      &       3.9560  &       0.0110  &       $-0.1340$       &       0.0050  &       0.0260  \\
K04     &       Cep     &       1.1     &       2.4     &       9       &       3.9530  &       0.0006  &       $-0.1336$       &       0.0008  &       0.0150  \\
vB99    &       I, II, III      &       2       &       8       &       163     &       3.8862  &       0.0260  &       $-0.1115$       &       0.0050  &       0.0235  \\
G04     &       III     &       1.6     &       3.9     &       74      &       3.9287  &       0.0070  &       $-0.1225$       &       0.0025  &       0.0023  \\
G04     &       I, II   &       -0.85   &       4.1     &       21      &       3.9172  &       0.0095  &       $-0.1215$       &       0.0040  &       0.0024  \\
P19     &       III     &       2.07    &       2.71    &       41      &       3.9537  &       0.0017  &       $-0.1330$       &       0.0017  &       0.0018  \\
S21     &       II, III &       1.8     &       3.8     &       70      &       3.9278  &       0.0016  &       $-0.1220$       &       0.0006  &       0.0022  \\
\hline
&               &       mag     &       mag     &               &       mag     &       mag     &               &               &       mag     \\
                        \hline
                \end{tabular}
                        \end{center}
{Notes: The references are the following:  dB93 \citep{dibenedetto93}, F97 \citep{fouque97}, N02 \citep{nordgren02}, K04 \citep{kervella04c}, vB99 \citep{vanbelle99}, G04 \citep{gro04b}, 
P19 \citep{gp13}, S21 \citep{salsi21}.  See appendix F of \cite{nardetto18} for more details about these SBCRs. The second, third, and fourth columns provide the validity domain in terms of class and $(V-K)_0$ color. Only SBCRs dB93 (III) and dB93 (I, II) are not corrected for extinction (we kept them for comparison). $N$ is the number of measurements over which the calibration was performed. $b$ and  $a$ are the coefficients of the SBCR in the form $F_{V_{0}} = a  (V-K)_0 + b  $. The statistical uncertainty on the coefficients of the SBCR is also provided, as is the rms. The SBCRs are based on various definitions in the literature. Some transformations were necessary to derive the coefficients of the SBCRs, their uncertainties, and the rms.}
\end{table*}

\setlength{\tabcolsep}{0.21cm}
\begin{table*}
\caption{Coefficients $a_k$ for polynomial SBCRs in the form $S_{V_{0}} = \sum \limits_{k=0}^{k} a_k (V-K)_0^k$ with their corresponding uncertainty $\sigma_\mathrm{a_k}$.  $l_1=\mathrm{(V-K)}_1$ and $l_2=\mathrm{(V-K)}_2$ are the limit of the validity domain of the SBCRs. The number of measurements N over which the calibrations of the five SBCRs listed in this table  were performed are 14, 132, 12, 41, and 41. The rms of this table (based on the $S_{V_{0}}$ definition of the surface brightness) should be divided by 10 (following Eq. \ref{Eq_conv}) to be compared to the rms of Table~\ref{tab_lin_sbcr}, which is based on the $F_{V_{0}}$ definition of the SBCR.}\label{tab_nlin_sbcr}
        \begin{center}
                \begin{tabular}{cccccccccccc}
                        \hline
                        \hline
                        ref & class & $l_1$     &       $l_2$   &$a_0 {_{\pm \sigma_\mathrm{a_0}}} $ & $ a_1 {_{\pm \sigma_\mathrm{a_1}}}  $ & $ a_2 {_{\pm \sigma_\mathrm{a_2}}} $ & $ a_3 {_{\pm \sigma_\mathrm{a_3}}} $ & $ a_4 {_{\pm \sigma_\mathrm{a_4}}} $ & $ a_5 {_{\pm \sigma_\mathrm{a_5}}} $ & rms \\
                        \hline
                         dB98  & III & 1.5 & 3.7   &$2.657$  & $1.421$  & $-0.033$  &   &   & & 0.03 \\
                         C14 & all & -0.9 & 3.7  & $2.624_{\pm 0.009}$ & $1.798_{\pm 0.020}$ & $-0.776_{\pm 0.034}$ & $0.517_{\pm 0.036}$ &  $-0.150_{\pm 0.015}$ &  $0.015_{\pm 0.002}$  & 0.10 \\
                         C14 & I+II & -0.88 & 3.21  & 2.291 & 2.151 & -0.461  & 0.073  &  &  & 0.08 \\
                         C14 & III & -0.74 & 3.69  & 2.497 & 1.916 & -0.335  & 0.050  &  &  & 0.07 \\
                         dB05 & all & -0.1 & 3.7  &$2.565_{\pm 0.016}$  & $1.483_{\pm 0.015}$ & $-0.044_{\pm 0.005}$ & &   &  &0.04 \\
                        \hline
                \end{tabular}
        \end{center}
{ Notes: The references are the following:  dB98 \citep{dibenedetto98} [Eq. 3], C14 \citep{challouf14}, dB05 \citep{dibenedetto05}.}
\end{table*}

\begin{table}
\begin{center}
\caption{ List of the infrared photometric systems we used to calibrate the linear and nonlinear SBCRs indicated in Tables \ref{tab_lin_sbcr} and \ref{tab_nlin_sbcr}, respectively. }\label{tab_systems}
\begin{tabular}{lcc}
\hline \hline \noalign{\smallskip}
Ref & class & photometric systems   \\ 
\hline
dB93    &       III     &       Johnson \\
dB93    &       I, II   &       Johnson \\
dB93    &       I, II   &       Johnson \\
F97     &       III     &       Johnson \\
F97     &       I, II   &       Johnson \\
F97     &       Cep     &       Johnson \\
N02     &       III     &       Johnson \\
N02     &       Cep     &       Johnson + CTIO  \\
K04     &       Cep     &       SAAO    \\
vB99    &       I, II, III      &       various \\
G04     &       III     &       Johnson + IRC   \\
G04     &       I, II   &       Johnson + IRC   \\
P19     &       III     &       SAAO    \\
S21     &       II, III &       Johnson + IRC + SAAO    \\
\hline
dB98  & III &  Johnson \\
C14 & all &  Johnson \\
C14 & I+II & Johnson  \\
C14 & III &  Johnson \\
dB05 & all &  Johnson \\
\hline \noalign{\smallskip}
\end{tabular}
\end{center}
\end{table}

\section{Application of the inverse BW method}\label{s_BW}


We derived the limb-darkened angular diameters (Eq.~\ref{Eq_Fv}) together with the associated uncertainties 
(Eq.~\ref{Eq_coe} and \ref{Eq_pho})
at the specific phases of $K$ band measurements (Fig.~\ref{fig_photo} bottom panel). The $V$ measurements were interpolated at the corresponding phases. 

\begin{figure}[htbp]
\begin{center}
\resizebox{0.8\hsize}{!}{\includegraphics[clip=true]{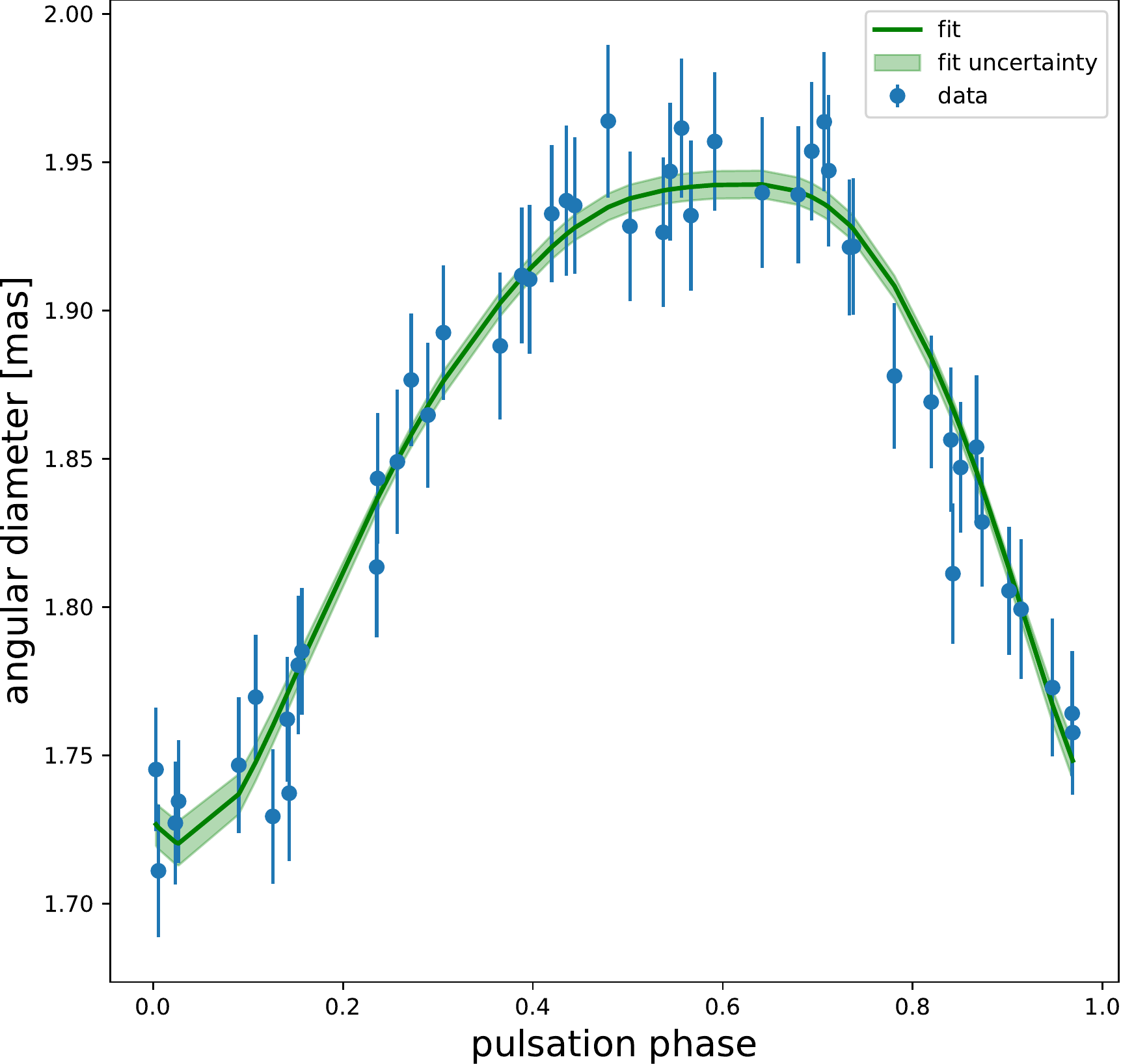}}
\end{center}
\caption{Example of the fit when the inverse BW method was applied, using the K04 SBCR together with the V- and K-band photometry of Fig.~\ref{fig_photo}. The derived angular diameters are plotted as a function of the pulsation phase together with the uncertainties propagated from the uncertainties on the $V$ and $K$ photometries. The best fit is indicated by the green line.}
\label{fig_bw}
\end{figure}

\begin{figure*}[htbp]
\begin{center}
\resizebox{0.8\hsize}{!}{\includegraphics[clip=true]{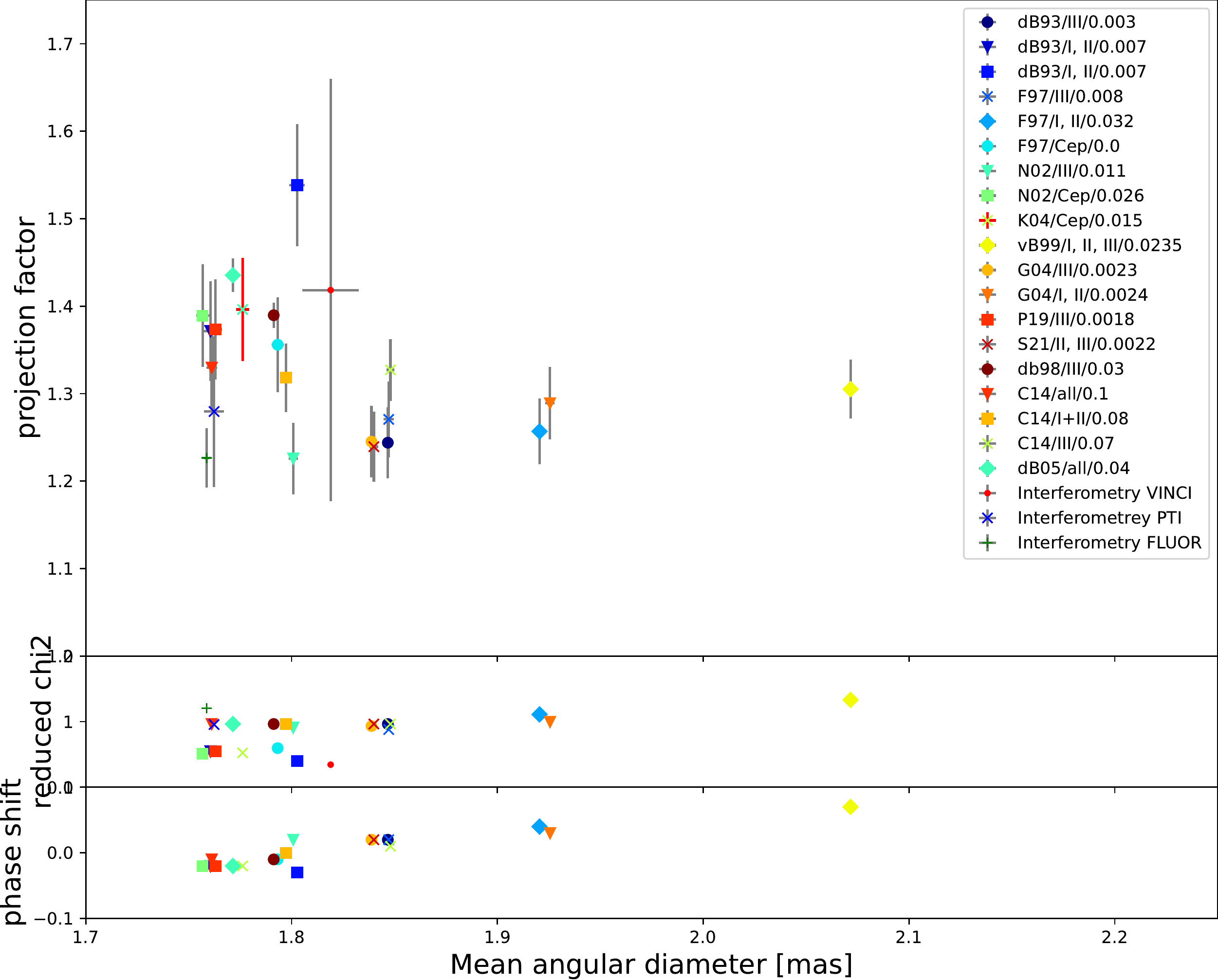}}
\end{center}
\caption{Top: Dependence of the calculated projection factors on the mean angular diameters from our analysis (from Table~\ref{tab_res} ) when the stellar parallax from \citet{benedict22} was adopted. Values calculated using both the different SBCRs and the  interferometric angular diameter curves are presented. Middle: Reduced $\chi^2$  plotted as a function of the mean angular diameter. Bottom: Phase shift plotted as a function of mean angular diameter.}\label{fig_bw_p}
\end{figure*}

To do this, we applied the ${\chi}^{2} $ minimization, 

\begin{equation}
\label{chi2sum} {\chi}^{2} = \sum_{i}{\frac{(\theta_{\rm
obs}(\phi_{i}) - \theta_{\rm model}(\phi_{i}))^2}{\sigma_{\rm
obs}(\phi_{i})^2}},
\end{equation}

where

$ \theta_{\rm obs}(\phi_{i})$ are the limb-darkened angular diameters derived either from interferometry or from the different SBCRs (Eq.~\ref{Eq_Fv}). Here, $\phi_{i}$ is the pulsation phase corresponding to the $i$th measurement;
$\sigma_{\rm obs}(\phi_{i})$ are the statistical uncertainties corresponding to interferometric measurements or the uncertainties associated with the SBCR as defined in 
Eq.~\ref{Eq_coe} and \ref{Eq_pho}. 
In the following, we consider either one of these uncertainties or the quadratic sum of all of them depending on the purpose;
$\theta_{\rm model}(\phi_{i})$ are the modeled limb-darkened angular diameters, defined as
\begin{equation}\label{diam_mod}
\theta_{\rm model}(\phi_{i}) = \overline{\theta} +
9.3009\, p \pi \int RV(\phi_{i})\mathrm{d}\phi_{i} [{\rm mas}],
\end{equation}
where the conversion factor 9.3009 was defined using the solar radius given in \citet{prsa16}. In the following, $RV$ is the interpolated radial velocity curve: either $RV_\mathrm{cc-c}$, $RV_\mathrm{cc-g}$ or $RV_\mathrm{c}$ for the individual lines listed in Table~\ref{tab_lines}.  $\pi$ is the parallax of the star.
Recently, \citet{benedict22} reanalyzed the Hubble Space Telescope Fine Guidance Sensor (FGS) astrometry, together with reference star parallax and proper motion priors from Gaia EDR3 \citep{edr3}, in order to search for the close companion of $\eta$ Aql. In addition to the Cepheid, the $\eta$ Aql system is probably composed of an F-type star at 0.66 arcsec \citep{evans13, gallenne14b} and a closer companion with a spectral type of B9.8V \citep{evans91}.  We considered their best parallax (their table 13 with Gaia priors), $\pi = 3.71 \pm 0.07$ mas. As discussed in their paper, this value provides an absolute K-band magnitude consistent with the Leavitt law, and in our case, it provides realistic projection factors (otherwise lower than 1 when the values in their tables 11 or 12 are used, i.e., without Gaia priors). Similarly to \citet{benedict22}, we do not find irregularities in our radial velocity measurements. This suggests either a nearly face-on orientation or a very long period for the closer companion. The relative precision of this parallax is about 2\%.  This linearly translates into a statistical uncertainty on the projection factor.  The mean angular diameter $\overline{\theta}$ and the projection factor $p$ were fit in order to minimize ${\chi}^{2}$. Interestingly, Fig.~\ref{fig_diam} shows that the angular diameter curves associated with the different SBCRs follow the shape of the $V-K$ color variation, but with a slight shift in phase. In the minimization process, a phase shift between the radial velocity and the angular diameter was considered. The phase shift providing the lowest reduced $\chi^2$ was adopted.

Figure \ref{fig_bw} shows an example fit in the case of the K04 SBCR  and using $RV_\mathrm{cc-c}$. The uncertainties on the limb-darkened angular diameter correspond to the uncertainties associated with the photometry (Eq. \ref{Eq_pho}). We find $\theta_0=1.776 \pm 0.003$ mas and $p=1.40 \pm 0.06,$ with a reduced $\chi^2$ of 0.6. The phase shift for the radial velocity curve is $\phi_0=-0.02$ (see Table~\ref{tab_res}). 
In the next section, we quantify the impact of the SBCRs, the method we used to derive the radial velocity, the extinction, and possibly the CSE, on the projection factor.

\section{Analysis of the uncertainties associated with the BW projection factor}\label{s_impact}

\begin{figure*}[htbp]
\begin{center}
\resizebox{0.4\hsize}{!}{\includegraphics[clip=true]{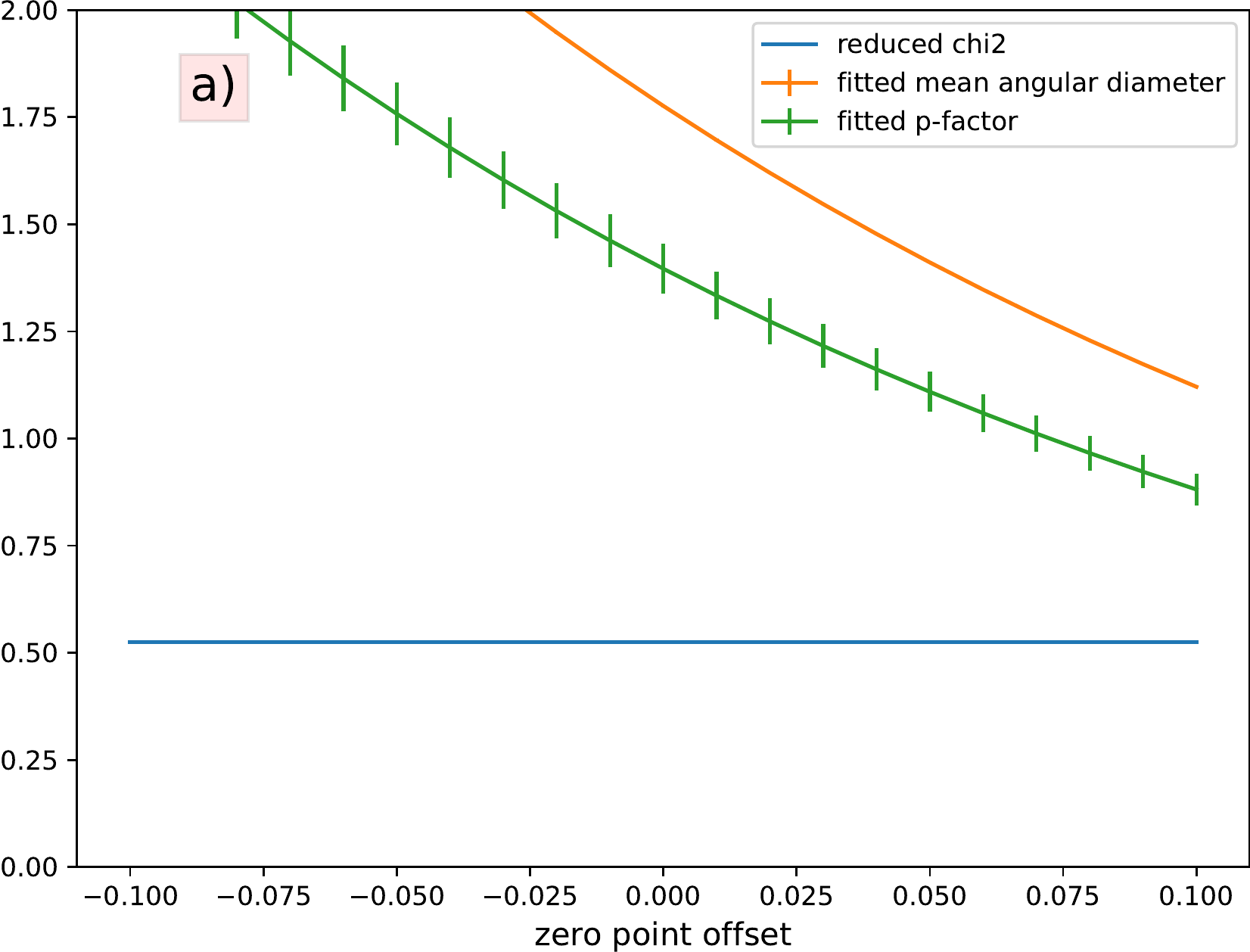}}
\resizebox{0.4\hsize}{!}{\includegraphics[clip=true]{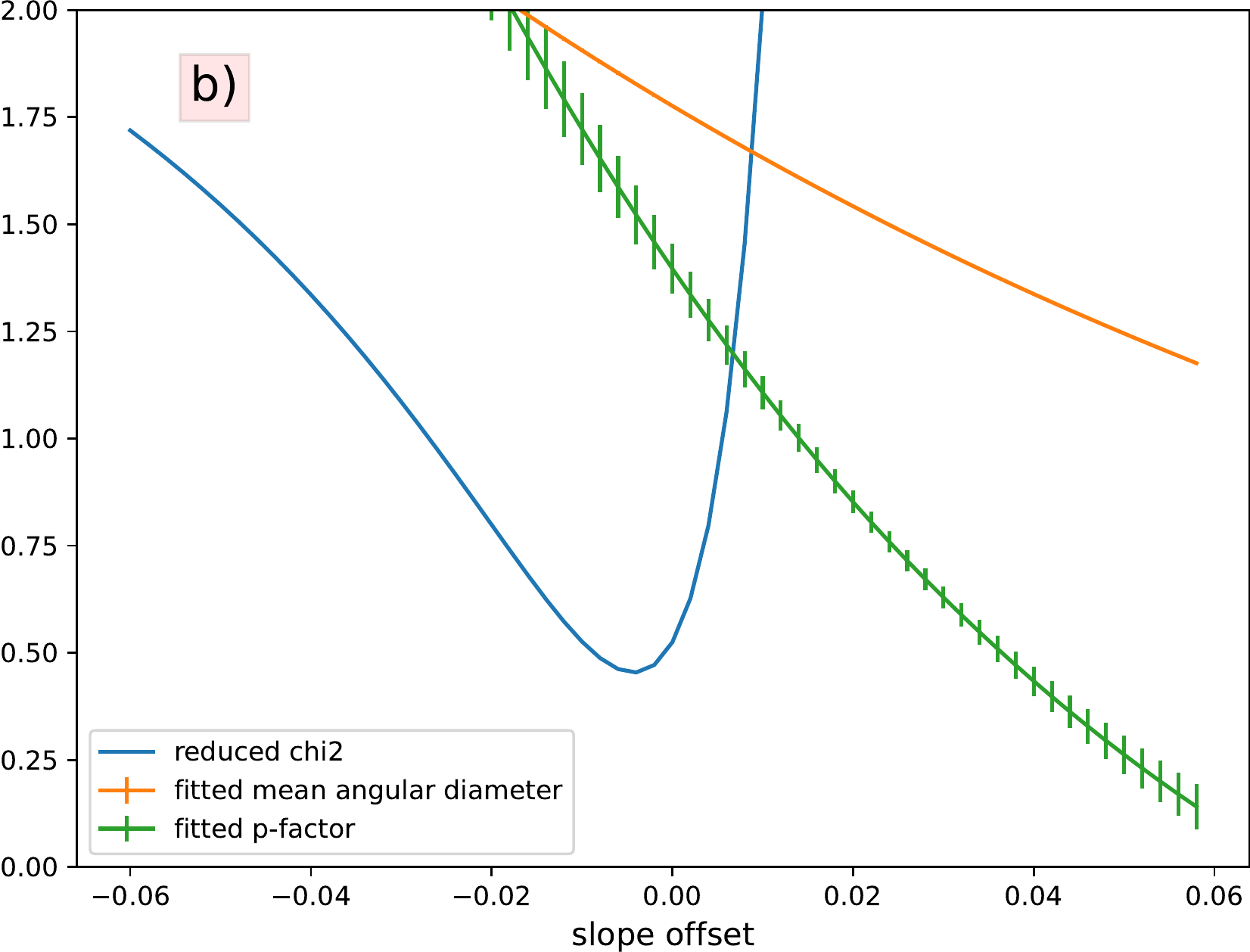}}
\end{center}
\begin{center}
\resizebox{0.4\hsize}{!}{\includegraphics[clip=true]{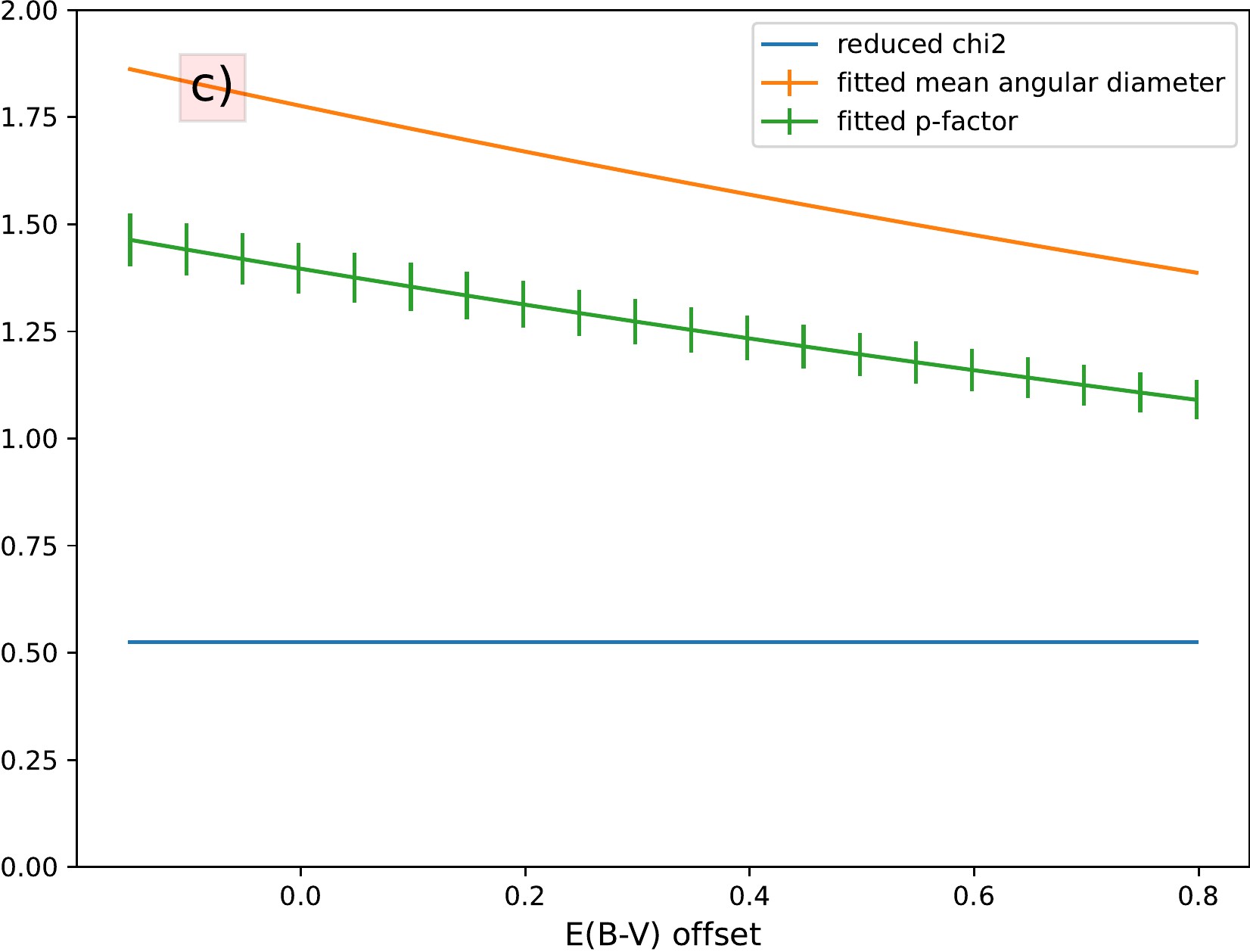}}
\resizebox{0.4\hsize}{!}{\includegraphics[clip=true]{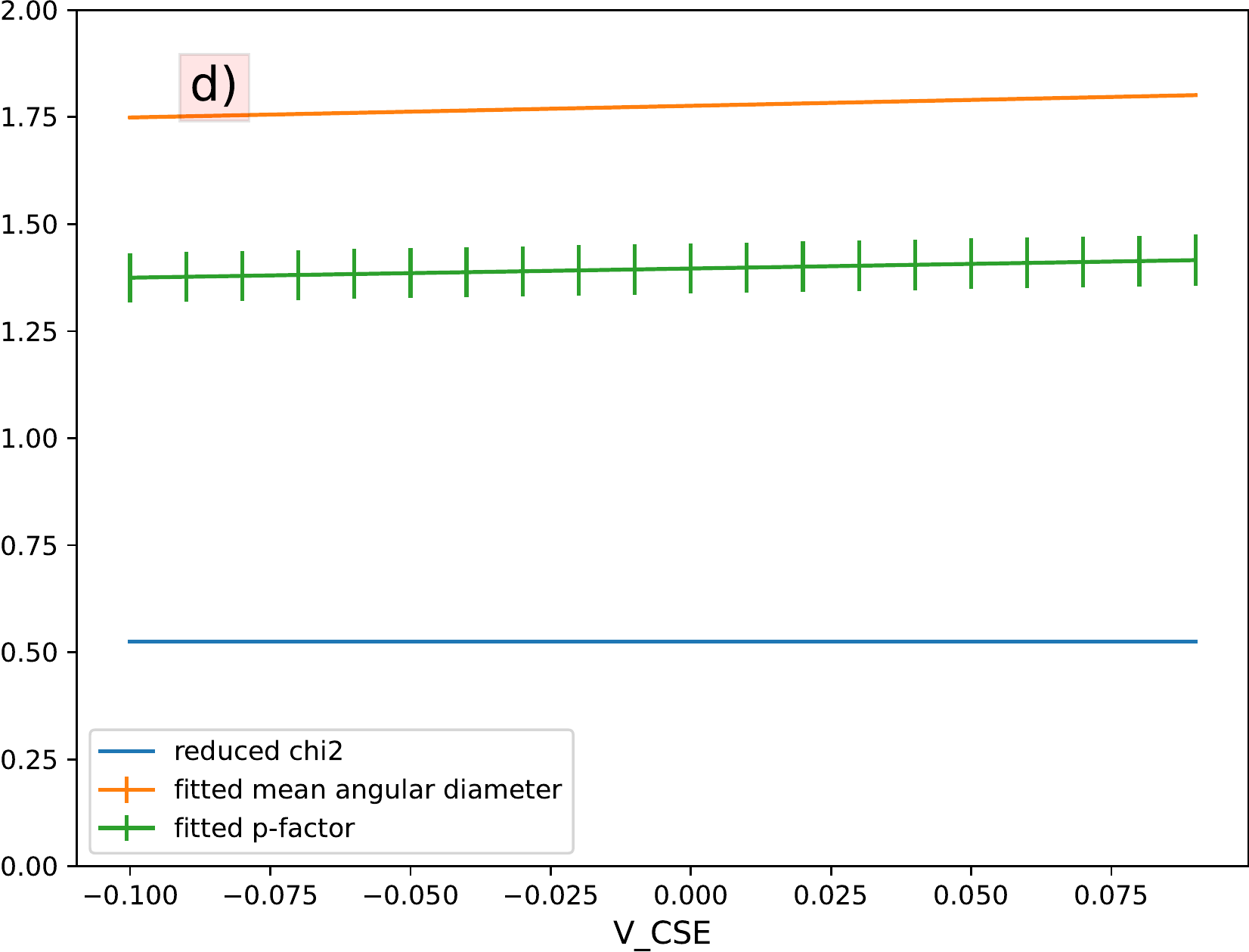}}
\end{center}
\begin{center}
\resizebox{0.4\hsize}{!}{\includegraphics[clip=true]{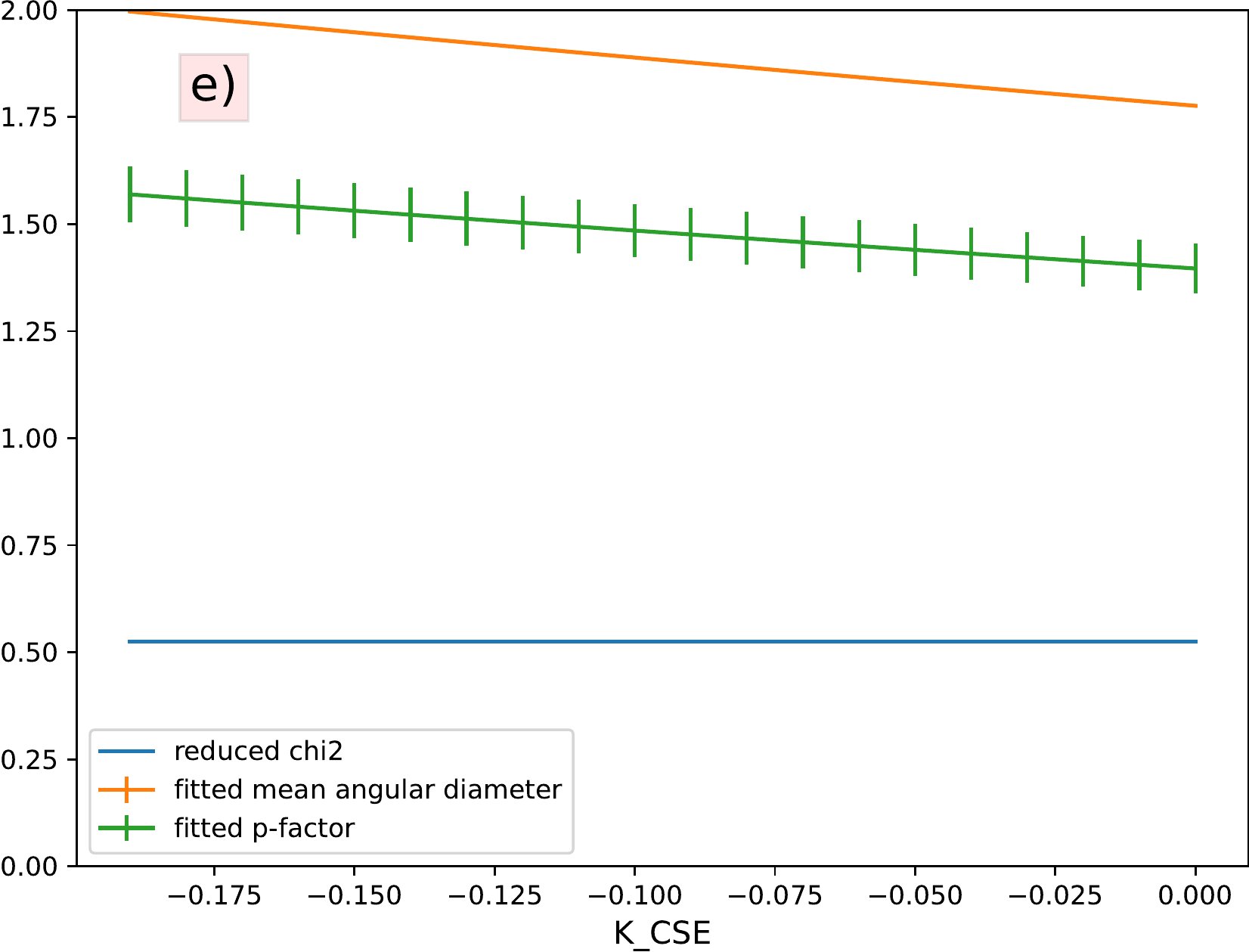}}
\end{center}
\caption{Application of the inverse BW method to $\eta$ Aql using the K04 SBCR. The derived mean angular diameter, the projection factor and  the reduced $\chi^2$ are shown when varying different parameters: (a) the zero point of the SBCR, (b) the slope of the SBCR, (c) the $ E(B-V)$, value (d) considering an offset on the V magnitude due to a CSE,  (e) considering an offset on the K magnitude due to a CSE. }
\label{fig_test}
\end{figure*}

We investigated different aspects, listed below, that might impact the BW projection factor.

\subsection{Uncertainty on the photometry}

This
is the first important aspect. If the uncertainties on the photometry are considered (Eq. \ref{Eq_pho}), they have a non-negligible impact on the projection factor. The mean uncertainty of measurements in the V and K band are 0.04 mag and  0.01 mag, respectively~(Fig.~\ref{fig_photo}).  When an uncertainty on  $A_{V}$ of 0.1 mag is also included, this translates into a mean uncertainty on the angular diameters of about 0.03 mas (Fig. \ref{fig_bw}), and finally, into a statistical uncertainty on the projection factor of 0.06 in the case of the K04 SBCR, that is, about 4\% (see Table~\ref{tab_res}). 

\subsection{Choice of the SBCR}

In the literature, the SBCRs have been calibrated using different samples of stars and interferometric instruments, which explains the significant difference obtained in Fig.~\ref{fig_sbcr}. The choice of the SBCR used to apply the inverse BW method is thus critical. We applied the inverse BW method (see previous section) for the different SBCRs of Tables~\ref{tab_lin_sbcr} and \ref{tab_nlin_sbcr} in order to derive the mean angular diameter and the projection factor. This was also done using interferometric measurements obtained with VINCI, PTI, and FLUOR. In Fig.~\ref{fig_bw_p} we plot the derived projection factors as a function of the mean angular diameter $\theta_0$. The uncertainties associated with the measurements correspond to Eq.~\ref{Eq_pho}, that is, the photometric uncertainties. The green cross with an error bar in red corresponds to the case of K04, which is our reference. In the second panel, we provide the reduced $\chi^2$, and the third panel shows the phase shift of the best fit. The results are listed in Table~\ref{tab_res}. The values of the projection factor range from 1.23 (N02 III) to 1.70 (vB99 I, II, III) with a statistical dispersion of 0.11 on the projection factor (or 8\%). The rms is an indicator of the robustness of the relations. When we exclude F97 (Cep), which provides no rms, the rms of SBCRs in Tables~\ref{tab_lin_sbcr} and \ref{tab_nlin_sbcr} ranges (using the $F_\mathrm{V}$ definition) from 0.0018 mag (P19) to 0.032 (F97 I, II) with a mean of 0.01 mag. The 5 SBCRs with the lowest rms are dB93 (III), G04 (III), G04 (I, II), P19 (III), and S21 (II, III) with 0.0030, 0.0023, 0.0024, 0.0018, and 0.0022 mag, respectively. The
$V-K$ validity domain of none of these relations is consistent with the color range of $\eta$~Aql, except for  G04 (I, II) (see Table~\ref{tab_lin_sbcr}). The K04 relation that is often used in the literature for the BW application has an rms of 0.015 mag, which is larger by an order of magnitude. 
The corresponding projection factors associated with these relations are 1.24$\pm 0.04$, 1.25$\pm 0.17$, 1.29$\pm 0.25$, 1.37$\pm 0.08,$ and 1.24$\pm 0.06$, respectively. In addition, when we consider the FLUOR/CHARA interferometric measurements, which have the best quality in term of precision and phase coverage, we find a projection factor of 1.23$\pm 0.03$. The upper panel of Fig.~\ref{fig_diam} shows that the K04 SBCRs provide an excellent agreement at maximum radius, but not at minimum radius (see phase 0.05), which leads to an overestimation of the amplitude of the angular diameter (compared to FLUOR) and thus to a larger projection factor of $1.40\pm 0.06$. This is the same for P19 (III)  (see Fig.~\ref{fig_diam}, and also Fig.~\ref{fig_bw_p}). We finally have four SBCRs
whose rms is lower than 0.03 mag and that provide consistent projection factors ($1.24-1.29$) that are also consistent with the projection factor derived from the best available interferometric data (FLUOR/CHARA, $p=1.23$). This analysis suggests that the projection factor of $\eta$ Aql is most probably about $1.25$. For the summary plot at the end of this section, we used the recent SBCR of S21 (II, III). 

\subsection{Uncertainties on the coefficients of the SBCR}

The statistical dispersion on the projection factor that is due to the uncertainty on the coefficients (Eq. \ref{Eq_coe}) of the K04 SBCR is 0.02 (or 1.5\%) when the K04 SBCR (Tab~\ref{tab_res}) is considered.

\subsection{Method for deriving the radial velocity}

As indicated in Sect. \ref{s_BW}, when we consider the K04 SBCR and the $\mathrm{RV_\mathrm{cc-c}}$ radial velocity (see Sect. \ref{s_HARPS-N}), we obtain $p=1.40 \pm 0.06$. When we instead use $\mathrm{RV_\mathrm{cc-g}}$, as is usually done in the literature, we obtain $p=1.33 \pm 0.06$, which is a non-negligible difference of 5\%. This is also shown in Fig.~\ref{fig_RVcurves1}, which shows that the amplitude of the $\mathrm{RV_\mathrm{cc-g}}$ is larger than the amplitude of the $\mathrm{RV_\mathrm{cc-c}}$ curve. As already discussed in \cite{nardetto06a}, the projection factor cannot be separated from the method that is used to derive the radial velocity. The impact of the choice of the spectral line can also be considered, instead of using thousands of lines, as in the cc method. In the 17 spectral lines of Table~\ref{tab_lines}, the projection factors range from 1.33 to 1.46 (with a standard deviation of 0.03). These values correspond to a range on the projection factor of 9\%. These results are consistent with \citet{nardetto07}. 

\subsection{Zeropoint of the SBCR}

In Fig. \ref{fig_test}a we quantify the impact of the zeropoint of the SBCR on the mean angular diameter and on the projection factor. To do this, we applied the inverse BW method and again considered the K04 SBCR as a reference ($p=1.40$). We considered only the statistical uncertainties associated with the photometry in the fit. The zeropoint of the SBCR was set to values between -0.1 (compared to the zeropoint of the  K04 SBCR) and 0.1, with steps of 0.01, which corresponds to the typical uncertainty on the zeropoint of SBCRs in Table~\ref{tab_lin_sbcr}. The resulting mean angular diameters, projection factors, and reduced $\chi^2$ are plotted as a function of the zeropoint offset. We find that an offset of $-0.01$ mag (+0.01 mag) on the zeropoint of the SBCR increases (decreases) the projection factor to $p=1.47$ ($p=1.33)$, which corresponds to $\pm$ 5\%. 

Table~\ref{tab_systems} shows that the SBCRs are based on various photometric systems, in particular in the infrared domain. An offset in photometry (in the visual and/or in the infrared) translates into an offset on the zeropoint of the SBCR, which is typically an order of magnitude lower in magnitude. Conversion relations between the various systems indicated in Table~\ref{tab_systems} in the literature (see for instance \cite{fouque97, gro04, glass85}) show that the offsets between the magnitudes are typically about 0.01 mag. This translates into an offset on the zeropoint of the SBCR of typically 0.001 mag, which corresponds to an offset on the projection factor of 0.5\%. Interestingly, the offset between the K-band magnitudes (regardless of the system considered) and the 2MASS system is larger, typically 0.044 magnitude \citep{bessell05}, which corresponds to an offset on the zeropoint of the SBCR of about 0.005 mag (see Table 5 of \citet{salsi21}) and finally to an non-negligible effect on the projection factor of about 2.5\%.

\subsection{Slope of the SBCR}

With the same method as for the zeropoint, we quantified the impact of the slope of the SBCR.  We thus considered different slopes from $-0.06$ to 0.06 with steps of 0.002 (which is the typical uncertainty on the slope of SBCR in Table~\ref{tab_lin_sbcr}), and we plot the resulting mean angular diameter, projection factor, and reduced $\chi^2$ in Fig. \ref{fig_test}b. We find that a difference $-0.002$ (0.002) on the slope of the SBCR increases (decreases) the projection factor to $1.46$ (1.33).
Interestingly, the reduced $\chi^2$ is highly sensitive to the slope of the SBCR. This suggests that the slope of the SBCR is critical for reproducing the shape to the angular diameter curve, that is, a shape that is consistent with the shape derived from the integrated radial velocity curve. It also shows that the projection factor is sensitive to the nonlinearity of the SBCRs listed in Table~\ref{tab_nlin_sbcr}.

\subsection{$ E(B-V)$  value}

In Fig. \ref{fig_test}c we show the impact of the $ E(B-V)$  value on the projection factor. In our reference fit (K04 SBCR), the $ E(B-V)$  value was  $0.152$ (see Sect. \ref{s_diameter}). For our test, we set $ E(B-V)$  from 0 to 0.8 mag (with steps of 0.05). When we consider $ E(B-V)$ $=0.1$ ($ E(B-V)$ $=0.2$), we find $p=1.42$ ($p=1.38$). This means that basically, an increase of 0.1 in $ E(B-V)$  translates into a decrease in the projection factor of 3\%. This is small because we used the $V-K$ color.

\subsection{Impact of the CSE in the visible domain}

In the SPIPS analysis \citep{merand05, trahin21}, an ad hoc{\it } infrared excess law is necessary to fit the atmosphere models. This infrared excess is physically understood by the presence of a CSE, whose physical nature is still under analysis \citep{hocde21}. It is also not excluded that the CSE emits some light in the visible domain \citep{nardetto16a} or even absorbs it \citep{hocde20a}. Thus, if a CSE like this exists, it might bias the visible magnitude (positively or negatively), which we can parameterize in Eq.~\ref{Eq_Fv} and Eq.~\ref{Eq_lin_sbcr} by replacing $V$ by $V+V_\mathrm{cse}$. If the observed magnitude  $V_\star$ is replaced  by $V_\star+V_\mathrm{cse}$ with $V_\mathrm{cse}$ positive (negative), the $V_\star+V_\mathrm{cse}$ magnitude is larger (smaller), which corresponds to a fainter (brighter) observed object (star + CSE) because the CSE is absorbing (emitting) light.
We also recall that an offset in magnitude for the CSE corresponds to a flux that is a constant fraction of the flux of the star.
We considered $V_\mathrm{cse}$ from $-0.1$ to 0.1 magnitudes (with steps of 0.01), and we plot the derived mean angular diameter, projection factor, and reduced $\chi^2$ in Fig.~\ref{fig_test}d. We find that $V_\mathrm{cse}=-0.1$ ($V_\mathrm{cse}=+0.1$) reduces (increases) the projection factor from $p=1.40$ to  $p=1.38$ ($p=1.42$). The effect of the CSE in the visible domain is rather small, about 1.5\%. However, this is a first-order analysis as the CSE can also involve phase-dependent mechanisms that could alter the amplitude of the $V$ and $K$ magnitudes. Moreover, absorption in the visible comes very likely with re-emission in the infrared, so that the impacts of CSE in the visible and in the infrared are correlated. Dedicated models are necessary to study the impact of these effects on the projection factor in detail.

\subsection{Impact of the CSE in the infrared domain}

We repeated the analysis in the infrared domain. We thus replaced $K$ by $K+K_\mathrm{cse}$ in Eq.~\ref{Eq_Fv} and Eq.~\ref{Eq_lin_sbcr}. However, we considered only negative values for $K_\mathrm{cse}$ (CSE emitting light) because so far, only infrared excess was observed in Cepheids, either in the spectral energy distributions \citep{trahin21, gallenne21, gro20} or by interferometry \citep{merand06, merand07}. However, we recall that a compact envelope of ionized gas can also potentially generate absorption in the K band (\citet{hocde20a}, their figure 10), but this needs to be investigated further. In this work, we thus considered $K_\mathrm{cse}$ values from 0.0 to $-$0.2 magnitudes with steps of 0.01 (Fig. \ref{fig_test}e). We find that $K_\mathrm{cse}=-0.1$~mag corresponds to an increase in the projection factor from $p=1.40$ to  $p=1.49$ (about 6\%), which is clearly not negligible.

\begin{figure}[htbp]
\begin{center}
\resizebox{1.0\hsize}{!}{\includegraphics[clip=true]{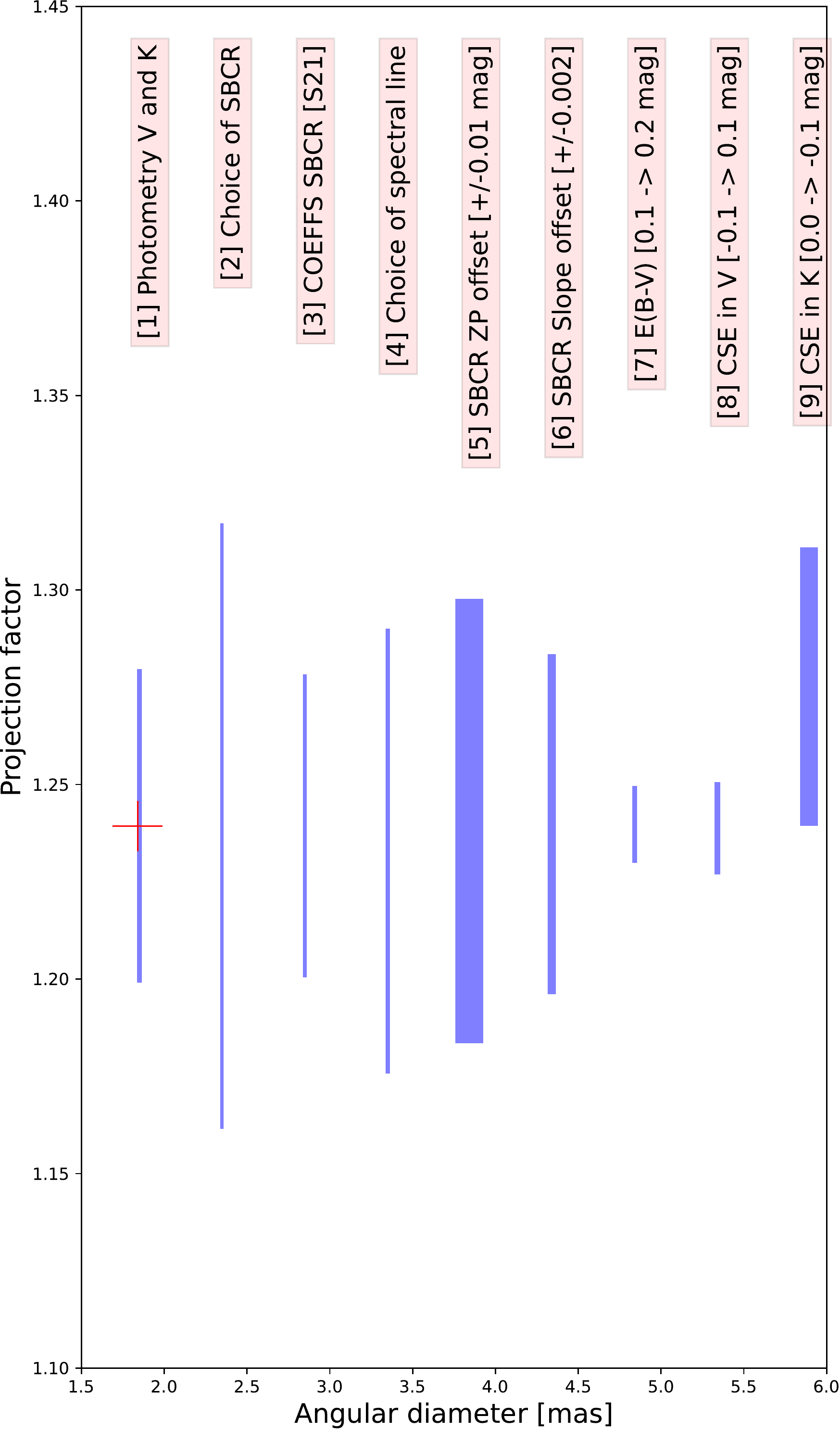}}
\caption{Different sources of statistical and systematical uncertainties that can affect the projection factor of Cepheids.
(1) Statistical uncertainty on the projection due to the uncertainty on the photometry ($V$, $K,$ and $A_\mathrm{V}$; blue bar). The resulting uncertainty on $\theta_\mathrm{0}$ is negligible (lower than the thickness of the bar). The red cross is the reference (see the text).
(2) Same as panel 1, except that the value of $\theta_\mathrm{0}$ is shifted by 0.5 mas for clarity. The blue bar corresponds to the systematical uncertainty on $p$ due to the choice of the SBCR in the literature, i.e., it corresponds to the standard deviation of $p$ values in Fig.~\ref{fig_bw_p} or in Table~\ref{tab_res}.
(3) Statistical uncertainties corresponding to the uncertainties on the coefficients of the S21 SBCR.  
(4) Range over which the projection factor varies (i.e., bias or systematics) depending on the choice of the spectral line in order to derive the radial velocity (among the list of 17 lines of Tab \ref{tab_lines}).   
(5) Indicative systematical uncertainties on $p$ and $\theta_\mathrm{0}$ due to a positive or negative 0.01 mag shift on the zeropoint of the SBCR. 
(6)  Indicative systematical uncertainties on $p$ and $\theta_\mathrm{0}$ due to a positive or negative 0.01 shift on the slope of the SBCR. 
(7) Offsets on $p$ and $\theta_\mathrm{0}$ when $ E(B-V)$  changes from $0.1$ to $0.2$ mag (instead of $0.152$).
(8) Offsets on $p$ and $\theta_\mathrm{0}$  when a shift in V from $-0.1$ to $0.1$ mag is considered (due to a possible CSE) instead of 0 in the original fit (item 1). 
(9) Offsets on $p$ and $\theta_\mathrm{0}$ when a shift in K of $-$0.1 mag is considered  (due to a possible CSE) instead of 0.
}\label{fig_summary}
\end{center}
\end{figure}

The nine uncertainty sources we discussed in this section are represented in the summary figure~\ref{fig_summary}.
In this figure, the uncertainties on the projection factor and the mean angular diameter of $\eta$~Aql are provided in different cases (shifted by $n*0.5$ mas for clarity, where $n$ is an integer). Following our discussion about the choice of the SBCR (section 7.2), we applied the inverse BW method to $\eta$ Aql using the S21 SBCR instead of K04. We considered the $RV_\mathrm{cc-c}$ definition of the velocity, the \citep{benedict22} parallax, and $ E(B-V) =0.152$ mag (see Sects. \ref{s_diameter} and \ref{s_BW} for details). The resulting projection factor is $1.24$ and $\theta_\mathrm{0}=1.84$ mas (red cross in the figure).  The different blue bars correspond to different cases, as indicated in the figure caption.



\begin{table*}
\begin{center}
\caption{Results of the inverse BW method when it was applied to different SBCRs (Cols. 1 and 2). The phase shift  $\phi_0$ and the   $\chi_\mathrm{red}$   are provided together with the results of the fit, i.e., the mean angular diameter   $\theta_\mathrm{0}$      and the projection factor $p$. The uncertainty on the projection factor is decomposed into uncertainties due to 
the uncertainty on the coefficients of the SBCR ($\sigma_{p_\mathrm{coe}}$, Eq.\ref{Eq_coe}) and to the uncertainty on the photometry ($\sigma_{p_\mathrm{pho}}$, Eq.\ref{Eq_pho}).  $\sigma_{p_\mathrm{tot}}$ is the quadratic sum of these uncertainties. 
For the nonlinear SBCRs of Table \ref{tab_nlin_sbcr}, only $\sigma_{p_\mathrm{coe}}$ is provided. 
For  interferometric results, i.e., VINCI \citep{kervella04a}, PTI \citep{lane02}, and FLUOR \citep{merand15}, only the total uncertainty on the projection factor due to the uncertainty on the angular diameter measurements is provided. In the first column, the plus and asterisk indicate that the $V-K$ range of $\eta$ Aql is not consistent or partially consistent, respectively, with the validity domain of the SBCR.
} \label{tab_res}
\setlength{\doublerulesep}{\arrayrulewidth}
 \tiny
\begin{tabular}{ccccccccccc}
\hline \hline \noalign{\smallskip}

REF     &   CLASS  &  $\phi_0$  &       $\chi_\mathrm{red}$      &       $\theta_\mathrm{0}$      &       $\sigma_{\theta_\mathrm{0}}$      &      $ p$         &       $\sigma_{p_\mathrm{coe}}$  &   $\sigma_{p_\mathrm{pho}}$   & $\sigma_{p_\mathrm{tot}}$ \\        
\hline
dB93$^{\ast}$  & III        & $+$0.02 & 1.0 & 1.847 & 0.002 & 1.24 & 0.00 & 0.04 & 0.04\\ 
dB93  & I, II      & $-$0.02 & 0.5 & 1.761 & 0.003 & 1.37 & 0.00 & 0.06 & 0.06\\ 
dB93  & I, II      & $-$0.03 & 0.4 & 1.803 & 0.004 & 1.54 & 0.00 & 0.07 & 0.07\\ 
F97$^{+}$   & III        & $+$0.02 & 0.9 & 1.847 & 0.002 & 1.27 & 0.29 & 0.04 & 0.29\\ 
F97   & I, II      & $+$0.04 & 1.1 & 1.920 & 0.002 & 1.26 & 0.57 & 0.04 & 0.57\\ 
F97   & Cep        & $-$0.01 & 0.6 & 1.793 & 0.003 & 1.36 & 0.00 & 0.05 & 0.05\\ 
N02   & III        & $+$0.02 & 0.9 & 1.801 & 0.002 & 1.23 & 0.12 & 0.04 & 0.13\\ 
N02   & Cep        &$-$0.02 & 0.5 & 1.757 & 0.003 & 1.39 & 0.26 & 0.06 & 0.27\\ 
K04   & Cep        &$ -$0.02 & 0.5 & 1.776 & 0.003 & 1.40 & 0.02 & 0.06 & 0.06\\ 
vB99  & Var        &$ +$0.08 & 1.3 & 2.684 & 0.002 & 1.69 & 1.94 & 0.04 & 1.95\\ 
vB99$^{+}$  & I, II, III & $+$0.07 & 1.3 & 2.072 & 0.002 & 1.31 & 0.67 & 0.03 & 0.67\\ 
G04$^{\ast}$   & III        & $+$0.02 & 0.9 & 1.839 & 0.002 & 1.25 & 0.17 & 0.04 & 0.17\\ 
G04   & I, II      & $+$0.03 & 1.0 & 1.926 & 0.002 & 1.29 & 0.25 & 0.04 & 0.25\\ 
P19$^{+}$   & III        & $-$0.02 & 0.5 & 1.763 & 0.003 & 1.37 & 0.05 & 0.06 & 0.08\\ 
S21$^{+}$   & II, III    & $+$0.02 & 1.0 & 1.840 & 0.002 & 1.24 & 0.04 & 0.04 & 0.06\\ 
db98$^{\ast}$  & III        & $-$0.01 & 0.1 & 1.791 & 0.001 & 1.39 & 0.01 &       &       &       \\ 
C14   & all        & $-$0.01 & 0.0 & 1.761 & 0.003 & 1.33 & 0.05 &       &       &       \\ 
C14   & I+II       & $+$0.00 & 0.0 & 1.797 & 0.002 & 1.32 & 0.04 &       &       &       \\ 
C14   & III        & $+$0.01 & 0.0 & 1.848 & 0.002 & 1.33 & 0.04 &       &       &       \\ 
dB05  & all        & $-$0.02 & 0.1 & 1.771 & 0.001 & 1.44 & 0.02 &       &       &       \\ 
VINCI &            & $+$0.01 & 0.3 & 1.819 & 0.014 & 1.42 &       &            & 0.24 \\ 
PTI   &            & $-$0.02 & 1.0 & 1.762 & 0.005 & 1.28 &       &            & 0.09 \\ 
FLUOR &            & $-$0.01 & 1.2 & 1.759 & 0.001 & 1.23 &       &             & 0.03 \\ 
\hline
     &    &    &  &  mas   &  mas     &    &    &   &   &  \\
\hline 
\end{tabular}
\end{center}
\end{table*}

\section{Conclusion}\label{s_conclusion}

The projection factor of $\eta$ Aql  clearly depends on the method that is used when the inverse BW method is applied: interferometry, SBCR, or even a combination of both (SPIPS). The choice of the SBCR is particularly crucial, as is the method that is used to derive the radial velocity curve: $\mathrm{RV_\mathrm{cc-g}}$, $\mathrm{RV_\mathrm{cc-c}}$ or even single-line analysis ($\mathrm{RV_\mathrm{c}}$). In addition to this, the projection factor might be biased by the calculation of extinction and by the presence or absence of a CSE. This CSE can indeed create an offset in the visible magnitude, in the infrared, or even in both. In addition, the statistical precision that can be expected on the projection is closely linked to the precision of the photometry, the precision on the coefficients of the SBCR, and also to the precision of the stellar parallax. 

Five SBCRs are suitable for Cepheids in the literature with an rms lower than 0.003 mag (using the $F_\mathrm{V}$ definition). Four of these five relations are consistent with each other within the $V-K$ validity domain of $\eta$ Aql. These four SBCRs also provide amplitudes of the angular diameter curve that are consistent with the best interferometric observations to date (FLUOR). However, offsets in terms of mean angular diameters (typically 0.1-0.2 mas) are found between the angular diameter curves from these SBCRs and the interferometric one. Thus, when these four SBCRs are considered together with the $\mathrm{RV_\mathrm{cc-c}}$ definition of the radial velocity, we find $p=1.24 \pm 0.04$ (dB93 III), $p=1.25 \pm 0.17$ (G04 III), $p=1.29 \pm 0.25$ (G04 I and II), and $p=1.24 \pm 0.06$ (S21 II and III), and $p=1.23 \pm 0.03$ when FLUOR interferometric measurements are considered. Including a 2\% statistical precision due to the uncertainty on the parallax  \citep{benedict22} does not change the global uncertainties on these projection factor. Thus, evidence indicates a projection factor for $\eta$ Aql of about 1.25, which agrees with hydrodynamical models \citep{nardetto04} and projection factors from eclipsing binaries \citep{pilecki13} for short-period Cepheids. Nevertheless, additional studies are necessary to confirm this result. In particular, the disagreements between SBCRs in the literature need clarification, even if the their origin is probably due to the use of different stellar samples, methods, and instruments. The impact of the CSE of Cepheids when calibrating and using the SBCRs needs clarification as well. In particular, the projection factors that we obtain (around 1.25) assume that $\eta$ Aql has no CSE that could alter the visual and/or the K-band magnitudes. It is still unclear whether the CSE could bias the V and/or K-band magnitudes, and if it does, by which amount. This is under investigation \citep{trahin21, hocde20a, gro20}.


Thus, in order to compare the projection factor of Cepheids, a consistent method is required. First, a dedicated SBCR should be used that was calibrated using homogeneous photometric and interferometric measurements. The same photometric system should be used when the SBCR is applied. However, to use such an SBCR, it might be necessary to verify the consistency of the SBCR for various Cepheids with different periods and at all pulsation phases. 
The upcoming instrument CHARA/SPICA \citep{mourard22} may help to improve the SBCR of Cepheids in the near future. Second, if possible, a single-line analysis (or a few lines) should be used instead of the cross-correlation approach. Importantly, the centroid method is strongly favored as it is independent of the rotation and the full width at half maximum of the line. Third, the CSE of Cepheids in the instability strip should be studied, characterized, and parameterized in the calibration or use of the SBCR. This can also be done in the SPIPS analysis. Fourth, it also seems important to secure homogeneous and precise photometric and spectroscopic data in order to improve the statistical precision on the projection. The community should achieve this in the coming years to understand the projection factor in depth, and thus to understand the physics of Cepheids.

\begin{acknowledgements}
The observations leading to these results have received funding  from the European Commission's Seventh Framework Programme (FP7/2013-2016)  under grant agreement number 312430 (OPTICON). The authors thank the GAPS observers F.~Borsa, L.~Di Fabrizio, R.~Fares, A.~Fiorenzano, P.~Giacobbe, J.~Maldonado, and G.~Scandariato. The authors thank the CHARA Array, which is funded by the National Science Foundation through NSF grants AST-0606958 and AST-0908253 and by Georgia State University through the College of Arts and Sciences, as well as the W. M. Keck Foundation. This research has made use of the SIMBAD and VIZIER\footnote{Available at http://cdsweb.u- strasbg.fr/} databases at CDS, Strasbourg (France), and of the electronic bibliography maintained by the NASA/ADS system.  WG gratefully acknowledges financial support for this work from the BASAL Centro de Astrofisica y Tecnologias Afines (CATA) PFB-06/2007, and from the Millenium Institute of Astrophysics (MAS) of the Iniciativa Cientifica Milenio del Ministerio de Economia, Fomento y Turismo de Chile, project IC120009. 
WG also acknowledges support from the ANID BASAL project ACE210002.
Support from the Polish National Sci-
ence Center grant MAESTRO 2012/06/A/ST9/00269 and DIR/WK/2018/09 grants of the Polish Minstry of Science and Higher Education is also acknowledged. EP and MR acknowledge financial support from PRIN INAF-2014. 
The authors acknowledge the support of the French Agence Nationale de la Recherche (ANR), under grant ANR-15-CE31-0012- 01 (project UnlockCepheids) and the financial support from ``Programme National de Physique Stellaire'' (PNPS) of CNRS/INSU, France. A.G. acknowledges support from the ANID-ALMA fund No. ASTRO20-0059. B.P. gratefully acknowledges support from the Polish National Science Center grant SONATA BIS 2020/38/E/ST9/00486.  This work has made use of data from the European Space Agency (ESA) mission {\it Gaia}, processed by the {\it Gaia} Data Processing and Analysis Consortium (DPAC). Funding for the DPAC has been provided by national institutions, in particular the institutions participating in the {\it Gaia} Multilateral Agreement. The research leading to these results  has received funding from the European Research Council (ERC) under the European Union's Horizon 2020 research and innovation program (projects CepBin, grant agreement 695099, and UniverScale, grant agreement 951549). VH is supported by the National Science Center, Poland, Sonata BIS project 2018/30/E/ST9/00598

\end{acknowledgements}

\bibliographystyle{aa}  
\bibliography{bibtex_nn} 

\break
\begin{appendix}

\section{Cross-correlated radial velocity tables}

\begin{table*}
\begin{center}
\caption{HARPS-N RV$\mathrm{cc-g}$ (F6 template) radial velocities of $\delta$~Cep.
} \label{Tab_log_ccgF6}
\setlength{\doublerulesep}{\arrayrulewidth}
 \tiny
\begin{tabular}{|lcrl|lcrl|}
\hline \hline \noalign{\smallskip}

BJD     &       $\phi$  &       RV$_\mathrm{cc-g}$      &       $\sigma_\mathrm{RV_\mathrm{cc-g}}$      &       BJD     &       $\phi$  &       RV$_\mathrm{cc-g}$      &       $\sigma_\mathrm{RV_\mathrm{cc-g}}$      \\

\hline
2457108.7601 &     0.589 &  -15.2019 &    0.0006 & 2457172.7213 &     0.501 &  -16.1754 &    0.0007 \\
2457109.7545 &     0.727 &   -5.6266 &    0.0006 & 2457173.5244 &     0.613 &  -14.5359 &    0.0009 \\
2457112.7071 &     0.139 &  -31.6954 &    0.0014 & 2457173.5264 &     0.613 &  -14.5281 &    0.0009 \\
2457112.7421 &     0.143 &  -31.6830 &    0.0006 & 2457173.7199 &     0.640 &  -13.2138 &    0.0008 \\
2457113.7377 &     0.282 &  -27.0033 &    0.0004 & 2457173.7212 &     0.640 &  -13.2045 &    0.0008 \\
2457114.7387 &     0.422 &  -20.4370 &    0.0008 & 2457174.5091 &     0.750 &   -3.1508 &    0.0017 \\
2457114.7425 &     0.422 &  -20.4143 &    0.0014 & 2457174.5122 &     0.750 &   -3.0957 &    0.0014 \\
2457137.7424 &     0.627 &  -14.0235 &    0.0006 & 2457174.5881 &     0.761 &   -1.9746 &    0.0008 \\
2457137.7460 &     0.627 &  -14.0025 &    0.0011 & 2457174.5912 &     0.761 &   -1.9304 &    0.0008 \\
2457142.7262 &     0.321 &  -25.2060 &    0.0008 & 2457175.5492 &     0.895 &    8.7276 &    0.0018 \\
2457143.7003 &     0.457 &  -18.4493 &    0.0007 & 2457175.5505 &     0.895 &    8.7335 &    0.0019 \\
2457143.7013 &     0.457 &  -18.4394 &    0.0007 & 2457175.7155 &     0.918 &    8.4551 &    0.0012 \\
2457144.7062 &     0.597 &  -14.9161 &    0.0009 & 2457175.7167 &     0.918 &    8.4455 &    0.0012 \\
2457144.7072 &     0.597 &  -14.9119 &    0.0009 & 2457176.5409 &     0.033 &  -19.8876 &    0.0016 \\
2457145.7086 &     0.737 &   -4.6713 &    0.0006 & 2457176.5422 &     0.033 &  -19.9350 &    0.0017 \\
2457145.7099 &     0.737 &   -4.6544 &    0.0006 & 2457176.7191 &     0.058 &  -25.4257 &    0.0009 \\
2457146.6917 &     0.874 &    7.8032 &    0.0013 & 2457176.7201 &     0.058 &  -25.4519 &    0.0010 \\
2457146.6930 &     0.874 &    7.8130 &    0.0015 & 2457177.5233 &     0.170 &  -31.1503 &    0.0011 \\
2457147.7245 &     0.018 &  -15.6986 &    0.0011 & 2457177.5242 &     0.170 &  -31.1459 &    0.0011 \\
2457148.7015 &     0.154 &  -31.4050 &    0.0008 & 2457177.6044 &     0.181 &  -30.8769 &    0.0010 \\
2457148.7025 &     0.154 &  -31.3978 &    0.0009 & 2457177.6054 &     0.181 &  -30.8731 &    0.0011 \\
2457153.7377 &     0.856 &    6.8569 &    0.0015 & 2457178.5383 &     0.311 &  -25.5400 &    0.0010 \\
2457153.7387 &     0.856 &    6.8702 &    0.0016 & 2457178.5398 &     0.311 &  -25.5330 &    0.0009 \\
2457154.6528 &     0.983 &   -4.5716 &    0.0042 & 2457178.7135 &     0.336 &  -24.3724 &    0.0011 \\
2457154.6548 &     0.983 &   -4.6670 &    0.0041 & 2457178.7146 &     0.336 &  -24.3631 &    0.0011 \\
2457155.6393 &     0.121 &  -31.4154 &    0.0075 & 2457203.5423 &     0.795 &    1.6335 &    0.0009 \\
2457156.7401 &     0.274 &  -27.4204 &    0.0018 & 2457203.5468 &     0.796 &    1.7043 &    0.0006 \\
2457157.6888 &     0.406 &  -21.3496 &    0.0019 & 2457204.5028 &     0.929 &    7.8041 &    0.0018 \\
2457157.6908 &     0.406 &  -21.3370 &    0.0021 & 2457204.5038 &     0.929 &    7.7895 &    0.0018 \\
2457158.7308 &     0.551 &  -15.0527 &    0.0010 & 2457205.5538 &     0.075 &  -28.3115 &    0.0010 \\
2457159.7412 &     0.692 &   -9.0430 &    0.0010 & 2457205.5548 &     0.076 &  -28.3322 &    0.0010 \\
2457159.7422 &     0.692 &   -9.0269 &    0.0011 & 2457206.5012 &     0.207 &  -29.9243 &    0.0009 \\
2457169.5437 &     0.058 &  -25.5432 &    0.0014 & 2457206.5022 &     0.208 &  -29.9202 &    0.0009 \\
2457169.5447 &     0.058 &  -25.5688 &    0.0013 & 2457207.5407 &     0.352 &  -23.6536 &    0.0008 \\
2457169.6200 &     0.069 &  -27.3812 &    0.0014 & 2457207.5417 &     0.352 &  -23.6476 &    0.0007 \\
2457169.6210 &     0.069 &  -27.4017 &    0.0015 & 2457208.5677 &     0.495 &  -16.3139 &    0.0008 \\
2457170.5647 &     0.200 &  -30.1823 &    0.0013 & 2457208.5687 &     0.495 &  -16.3042 &    0.0009 \\
2457170.5655 &     0.200 &  -30.1788 &    0.0013 & 2457209.6511 &     0.646 &  -12.6662 &    0.0011 \\
2457170.7115 &     0.221 &  -29.4770 &    0.0014 & 2457209.6521 &     0.646 &  -12.6586 &    0.0012 \\
2457170.7121 &     0.221 &  -29.4755 &    0.0013 & 2457210.6167 &     0.781 &    0.1251 &    0.0013 \\
2457170.7141 &     0.221 &  -29.4616 &    0.0012 & 2457210.6177 &     0.781 &    0.1348 &    0.0012 \\
2457170.7149 &     0.221 &  -29.4542 &    0.0011 & 2457269.5627 &     0.994 &   -8.4791 &    0.0017 \\
2457171.5293 &     0.335 &  -24.4193 &    0.0016 & 2457269.5637 &     0.994 &   -8.5256 &    0.0017 \\
2457171.5317 &     0.335 &  -24.4048 &    0.0015 & 2457271.5496 &     0.271 &  -27.4435 &    0.0008 \\
2457171.6082 &     0.346 &  -23.9309 &    0.0007 & 2457271.5506 &     0.271 &  -27.4338 &    0.0008 \\
2457171.6103 &     0.346 &  -23.9160 &    0.0007 & 2457272.5884 &     0.416 &  -20.6674 &    0.0008 \\
2457172.5913 &     0.483 &  -17.1094 &    0.0006 & 2457272.5893 &     0.416 &  -20.6576 &    0.0008 \\
2457172.5930 &     0.483 &  -17.0983 &    0.0007 & 2457273.5519 &     0.550 &  -15.0334 &    0.0008 \\
2457172.7199 &     0.500 &  -16.1830 &    0.0007 & 2457273.5529 &     0.550 &  -15.0359 &    0.0009 \\
\hline
days     &    &  \kms   &  \kms   & days     &    &  \kms   &  \kms  \\
\hline 
\end{tabular}
\end{center}
\end{table*}

\begin{table*}
\begin{center}
\caption{HARPS-N RV$\mathrm{cc-c}$ (F6 template) radial velocities of $\delta$~Cep. 
} \label{Tab_log_cccF6}
\setlength{\doublerulesep}{\arrayrulewidth}
 \tiny
\begin{tabular}{|lcrl|lcrl|}
\hline \hline \noalign{\smallskip}

BJD     &       $\phi$  &       RV$_\mathrm{cc-c}$      &       $\sigma_\mathrm{RV_\mathrm{cc-c}}$      &       BJD     &       $\phi$  &       RV$_\mathrm{cc-c}$      &       $\sigma_\mathrm{RV_\mathrm{cc-c}}$      \\

\hline
2457108.7601 &     0.589 &  -15.3761 &    0.0006 & 2457172.7213 &     0.501 &  -16.4848 &    0.0007 \\
2457109.7545 &     0.727 &   -6.3331 &    0.0006 & 2457173.5244 &     0.613 &  -14.7596 &    0.0009 \\
2457112.7071 &     0.139 &  -30.9229 &    0.0014 & 2457173.5264 &     0.613 &  -14.7451 &    0.0009 \\
2457112.7421 &     0.143 &  -30.9397 &    0.0006 & 2457173.7199 &     0.640 &  -13.4911 &    0.0008 \\
2457113.7377 &     0.282 &  -25.8473 &    0.0004 & 2457173.7212 &     0.640 &  -13.4843 &    0.0008 \\
2457114.7387 &     0.422 &  -20.4125 &    0.0008 & 2457174.5091 &     0.750 &   -4.0068 &    0.0017 \\
2457114.7425 &     0.422 &  -20.3984 &    0.0014 & 2457174.5122 &     0.750 &   -3.9731 &    0.0014 \\
2457137.7424 &     0.627 &  -14.2505 &    0.0006 & 2457174.5881 &     0.761 &   -2.9180 &    0.0008 \\
2457137.7460 &     0.627 &  -14.2298 &    0.0011 & 2457174.5912 &     0.761 &   -2.9023 &    0.0008 \\
2457142.7262 &     0.321 &  -24.7608 &    0.0008 & 2457175.5492 &     0.895 &    7.4015 &    0.0018 \\
2457143.7003 &     0.457 &  -18.6297 &    0.0007 & 2457175.5505 &     0.895 &    7.4186 &    0.0019 \\
2457143.7013 &     0.457 &  -18.5950 &    0.0007 & 2457175.7155 &     0.918 &    7.2674 &    0.0012 \\
2457144.7062 &     0.597 &  -15.1453 &    0.0009 & 2457175.7167 &     0.918 &    7.2925 &    0.0012 \\
2457144.7072 &     0.597 &  -15.1173 &    0.0009 & 2457176.5409 &     0.033 &  -19.8247 &    0.0016 \\
2457145.7086 &     0.737 &   -5.4541 &    0.0006 & 2457176.5422 &     0.033 &  -19.8756 &    0.0017 \\
2457145.7099 &     0.737 &   -5.4392 &    0.0006 & 2457176.7191 &     0.058 &  -25.0661 &    0.0009 \\
2457146.6917 &     0.874 &    6.5466 &    0.0013 & 2457176.7201 &     0.058 &  -25.0954 &    0.0010 \\
2457146.6930 &     0.874 &    6.5514 &    0.0015 & 2457177.5233 &     0.170 &  -30.5110 &    0.0011 \\
2457147.7245 &     0.018 &  -15.8904 &    0.0011 & 2457177.5242 &     0.170 &  -30.5049 &    0.0011 \\
2457148.7015 &     0.154 &  -30.6799 &    0.0008 & 2457177.6044 &     0.181 &  -30.2474 &    0.0010 \\
2457148.7025 &     0.154 &  -30.6725 &    0.0009 & 2457177.6054 &     0.181 &  -30.2386 &    0.0011 \\
2457153.7377 &     0.856 &    5.5143 &    0.0015 & 2457178.5383 &     0.311 &  -25.1140 &    0.0010 \\
2457153.7387 &     0.856 &    5.5248 &    0.0016 & 2457178.5398 &     0.311 &  -25.1088 &    0.0009 \\
2457154.6528 &     0.983 &   -4.9432 &    0.0042 & 2457178.7135 &     0.336 &  -24.0095 &    0.0011 \\
2457154.6548 &     0.983 &   -5.0159 &    0.0041 & 2457178.7146 &     0.336 &  -24.0044 &    0.0011 \\
2457155.6393 &     0.121 &  -30.7510 &    0.0075 & 2457203.5423 &     0.795 &    0.5924 &    0.0009 \\
2457156.7401 &     0.274 &  -26.8474 &    0.0018 & 2457203.5468 &     0.796 &    0.6528 &    0.0006 \\
2457157.6888 &     0.406 &  -21.1738 &    0.0019 & 2457204.5028 &     0.929 &    6.8420 &    0.0018 \\
2457157.6908 &     0.406 &  -21.1627 &    0.0021 & 2457204.5038 &     0.929 &    6.8340 &    0.0018 \\
2457158.7308 &     0.551 &  -15.3880 &    0.0010 & 2457205.5538 &     0.075 &  -27.7975 &    0.0010 \\
2457159.7412 &     0.692 &   -9.5467 &    0.0010 & 2457205.5548 &     0.076 &  -27.8137 &    0.0010 \\
2457159.7422 &     0.692 &   -9.5392 &    0.0011 & 2457206.5012 &     0.207 &  -29.3577 &    0.0009 \\
2457169.5437 &     0.058 &  -25.1674 &    0.0014 & 2457206.5022 &     0.208 &  -29.3583 &    0.0009 \\
2457169.5447 &     0.058 &  -25.1881 &    0.0013 & 2457207.5407 &     0.352 &  -23.2804 &    0.0008 \\
2457169.6200 &     0.069 &  -26.8909 &    0.0014 & 2457207.5417 &     0.352 &  -23.2737 &    0.0007 \\
2457169.6210 &     0.069 &  -26.8949 &    0.0015 & 2457208.5677 &     0.495 &  -16.5356 &    0.0008 \\
2457170.5647 &     0.200 &  -29.5809 &    0.0013 & 2457208.5687 &     0.495 &  -16.5355 &    0.0009 \\
2457170.5655 &     0.200 &  -29.5861 &    0.0013 & 2457209.6511 &     0.646 &  -12.9839 &    0.0011 \\
2457170.7115 &     0.221 &  -28.9194 &    0.0014 & 2457209.6521 &     0.646 &  -12.9836 &    0.0012 \\
2457170.7121 &     0.221 &  -28.9326 &    0.0013 & 2457210.6167 &     0.781 &   -0.8593 &    0.0013 \\
2457170.7141 &     0.221 &  -28.9101 &    0.0012 & 2457210.6177 &     0.781 &   -0.8316 &    0.0012 \\
2457170.7149 &     0.221 &  -28.8955 &    0.0011 & 2457269.5627 &     0.994 &   -8.6815 &    0.0017 \\
2457171.5293 &     0.335 &  -24.0206 &    0.0016 & 2457269.5637 &     0.994 &   -8.7346 &    0.0017 \\
2457171.5317 &     0.335 &  -24.0052 &    0.0015 & 2457271.5496 &     0.271 &  -26.7610 &    0.0008 \\
2457171.6082 &     0.346 &  -23.5647 &    0.0007 & 2457271.5506 &     0.271 &  -26.7476 &    0.0008 \\
2457171.6103 &     0.346 &  -23.5457 &    0.0007 & 2457272.5884 &     0.416 &  -20.4835 &    0.0008 \\
2457172.5913 &     0.483 &  -17.3346 &    0.0006 & 2457272.5893 &     0.416 &  -20.4687 &    0.0008 \\
2457172.5930 &     0.483 &  -17.3298 &    0.0007 & 2457273.5519 &     0.550 &  -15.2903 &    0.0008 \\
2457172.7199 &     0.500 &  -16.4919 &    0.0007 & 2457273.5529 &     0.550 &  -15.2798 &    0.0009 \\
\hline
days     &    &  \kms   &  \kms   & days     &    &  \kms   &  \kms  \\
\hline 
\end{tabular}
\end{center}
\end{table*}

\end{appendix} 
\end{document}